\documentclass[aps,prd,reprint,amsmath,amssymb,nofootinbib, superscriptaddress]{revtex4-2}

\usepackage{amsmath,mathtools,physics}
\usepackage{natbib}
\usepackage{graphicx}
\usepackage{dcolumn}
\usepackage{hyperref}
\usepackage{xcolor}

\usepackage{listings}
\usepackage{multirow}

\usepackage[makeroom]{cancel}
\usepackage{pgf}

\usepackage{tikz}
\usetikzlibrary{backgrounds,calc}
\usetikzlibrary{shapes.misc}
\usepackage{multirow}

\newcommand{\lwp}{\texttt{LWP}}

\usepackage[font=small,labelfont=bf,justification=justified]{caption}
\usepackage[font=small,labelfont=bf,justification=justified]{subcaption}

\allowdisplaybreaks
\makeatletter

\definecolor{lime}{HTML}{A6CE39}
\DeclareRobustCommand{\orcidicon}{
	\begin{tikzpicture}
	\draw[lime, fill=lime] (0,0) 
	circle [radius=0.16] 
	node[white] {{\fontfamily{qag}\selectfont \tiny ID}};
	\draw[white, fill=white] (-0.0625,0.095) 
	circle [radius=0.007];
	\end{tikzpicture}
	\hspace{-2mm}
}
\foreach \x in {A, ..., Z}{\expandafter\xdef\csname orcid\x\endcsname{\noexpand\href{https://orcid.org/\csname orcidauthor\x\endcsname}{\noexpand\orcidicon}}}

\begin{document}

\title{Inferring the neutron star equation of state with nuclear-physics informed semiparametric models}

\author{\large{Sunny Ng}\orcidA{}{}}\email{sunnyng.sshn@gmail.com} 
\affiliation{California State University Fullerton, Fullerton, CA 92831, USA\relax}
\def\sn{\textcolor{blue}}

\author{\large{Isaac Legred}\orcidB{}}
\affiliation{California Institute of Technology, Pasadena, CA 91125, USA}
\def\il{\textcolor{orange}}

\author{\large{Lami Suleiman}\orcidC{}}
\affiliation{California State University Fullerton, Fullerton, CA 92831, USA\relax}
\affiliation{Deutsches Elektronen-Synchrotron DESY, Platanenallee 6, 15738 Zeuthen, Germany}
\affiliation{Deutsches Zentrum f\"ur Astrophysik (DZA), Postplatz 1, 02826 G\"orlitz, Germany}
\def\ls{\textcolor{purple}}

\author{\large{Philippe Landry}\orcidD{}}
\affiliation{Canadian Institute for Theoretical Astrophysics, University of Toronto, Toronto, Ontario M5S 3H8, Canada}
\affiliation{Perimeter Institute for Theoretical Physics, Waterloo, Ontario N2L 2Y5, Canada}
\def\pl{\textcolor{cyan}}

\author{\large{Lyla Traylor}}
\affiliation{California State University Fullerton, Fullerton, CA 92831, USA\relax}
\def\lt{\textcolor{green}}

\author{\large{Jocelyn Read}\orcidE{}}\email{jread@fullerton.edu} 
\affiliation{California State University Fullerton, Fullerton, CA 92831, USA\relax}
\affiliation{Perimeter Institute for Theoretical Physics, Waterloo, Ontario N2L 2Y5, Canada}
\def\jr{\textcolor{brown}}

\date{\today}

\begin{abstract}
    Over the past decade, an abundance of information from neutron-star observations, nuclear experiments and theory has transformed our efforts to elucidate the properties of dense matter. However, at high densities relevant to the cores of neutron stars, substantial uncertainty about the dense matter equation of state (EoS) remains. In this work, we present a semiparametric equation of state framework aimed at better integrating knowledge across these domains in astrophysical inference. We use a Meta-model and realistic crust at low densities, and Gaussian Process extensions at high densities. Comparisons between our semiparametric framework to fully nonparametric EoS representations show that imposing nuclear theoretical and experimental constraints through the Meta-model up to nuclear saturation density results in constraints on the pressure up to twice nuclear saturation density. We also show that our Gaussian Process trained on EoS models with nucleonic, hyperonic, and quark compositions extends the range of EoS explored at high density compared to a piecewise polytropic extension schema, under the requirements of causality of matter and of supporting the existence of heavy pulsars. We find that maximum TOV masses above $3.2 M_{\odot}$ can be supported by causal EoS compatible with nuclear constraints at low densities. We then combine information from existing observations of heavy pulsar masses, gravitational waves emitted from binary neutron star mergers, and X-ray pulse profile modeling of millisecond pulsars within a Bayesian inference scheme using our semiparametric EoS prior. With information from all public NICER pulsars (including PSR J0030$+$0451, PSR J0740$+$6620, PSR J0437-4715, and PSR J0614-3329), we find an astrophysically favored pressure at two times nuclear saturation density of $P(2\rho_{\rm nuc}) = 1.98^{+2.13}_{-1.08}\times10^{34}$ dyn/cm$^{2}$, a radius of a $1.4 M_{\odot}$ neutron star value of $R_{1.4} = 11.4^{+0.98}_{-0.60}$\;km, and $M_{\rm max} = 2.31_{-0.23}^{+0.35} M_{\odot}$ at the 90\% credible level. 
\end{abstract}

\maketitle

\section{Introduction}

Understanding the strong interaction of matter in regimes of low temperatures and high density is a major challenge in nuclear physics. 
Experimental measurements of nuclei inform nuclear matter across a range of densities \cite{Koehn_2025, Tsang_2024,Reed_2021,Roca-Maza:2018ujj}.  Rare isotope beam facilities such as FRIB are exploring symmetric and asymmetric matter up to nuclear saturation density \cite{Crawford_2024}, and heavy ion collisions inform our understanding of nuclear matter at increasing ranges of temperature, density, and isospin asymmetry \cite{Senger:2024fwc,Kumar_2024,Sorensen_2024}. High density matter can be described by nuclear theory frameworks with quantified uncertainties, such as Chiral Effective Field Theory ($\chi$EFT), but these begin to break down by twice nuclear saturation density \cite{Drischler_2021}. Extending nuclear theory to very high density relies on unknown or poorly constrained degrees of freedom, potentially including pions, kaons, hyperons and quarks \cite{Chatziioannou_2024}. 

Neutron stars (NSs) are extremely compact astrophysical objects with interiors composed of strongly interacting neutron-rich matter at relatively low temperatures. 
Their macroscopic properties are determined by the Equation of State (EoS) of dense matter in beta equilibrium and the theory of gravitation. Observations of NSs can therefore be used in combination with theoretical and experimental nuclear physics results as a unique probe of dense matter.

Astrophysical observations have begun to produce robust constraints on the nuclear EoS. Measurements relying on general relativity have provided neutron star properties in multiple ways.  Radio observations of galactic pulsars provide precise measurements of the heaviest NS masses through pulsar radio timing\cite{2010Natur.467.1081D,2013Sci...340..448A,2021ApJ...915L..12F,2020NatAs...4...72C}. The largest masses inferred from those measurements imply that the pressure of strongly interacting matter inside the core of the star must be sufficiently high to counterbalance the self-gravitation of massive stars. The LIGO-Virgo-KAGRA network \cite{LVK_network_2018,2015CQGra..32g4001L,2015CQGra..32b4001A,2021PTEP.2021eA101A} of gravitational-wave observatories detects gravitational waves emitted from merging compact objects. From binary neutron star (BNS) mergers, the mass and tidal deformability can simultaneously be extracted from the signal. Gravitational wave data for the loudest BNS merger detected to date, GW170817 \cite{Abbott_2017}, provided significant upper limits on the tidal deformability, a parameter that quantifies the strength of the influence of neutron-star matter on the gravitational-wave signal \cite{LIGOScientific:2018hze,LIGOScientific:2018cki}. This observational constraint leads to an upper bound on the EoSs pressure around twice nuclear saturation density \cite{LIGOScientific:2018cki,De:2018uhw,Annala:2017llu}. The Neutron Star Interior Composition ExploRer (NICER) X-ray telescope \cite{Gendreau_2016} makes simultaneous measurements of the mass and radius of a few millisecond pulsars from X-ray pulse profile modeling (see \cite{Miller_2019, Dittmann_2024, Vinciguerra_2024, Salmi_2024, Choudhury_2024, Mauviard_2025} and references therein). The X-ray pulse profile data of the first pulsars observed favored models with moderately large radii, leading to a preference for EoSs with higher pressures; however, recent NICER observations are consistent with smaller radii like those inferred from GWs \cite{Choudhury_2024, Mauviard_2025}.

In order to constrain the EoS with astrophysical information, a Bayesian inference scheme requires a model and prior distribution of possible EoSs. There exists in the literature a number of parametric EoS constructions used to build those distributions, such as piecewise polytropes ~\cite{Read_2009}, spectral representations ~\cite{Lindblom_2010}, the meta-model ~\cite{Margueron_2018}, and sound speed parametrizations ~\cite{Tews_2018, Greif_2019}. 
More recently, non-parametric representations based on Gaussian Process (GP) regression, see Ref.~\cite{Landry_2019, EssickLandryHolz_2020,Gorda2023,Mroczek:2023eff}, have provided methods for generating EoSs with larger variability in the pressure-energy density space. Unlike parametrizations which implicitly require a particular functional form for the EoS, the GP method allows for prior support across the full space of causal and thermodynamically stable EoSs. A non-parametric approach therefore can avoid potential biases which arise from \textit{a priori} excluding the true EoS via an inapt choice of functional form for the parametrization~\cite{Legred_2022}. In contrast to fully agnostic approaches, the meta-model \cite{Margueron_2018} utilizes nuclear empirical parameters in the construction of the EoS. Said parameters are informed by terrestrial laboratory experiments, such as heavy ion collisions \cite{Tsang_2024} in order to build nuclear-physics informed EoSs.

In this work, we build on the approach taken in Ref.~\cite{Davis_2024}, and present a semiparametric method to build physically realistic EoSs based on the meta-modeling approach constrained by nuclear theory and experiments at low density, and using Gaussian Process generated extensions above nuclear density. Our methodology is presented in Section~\ref{sec:methods}. In Section~\ref{sec:methods/eos}, we detail our construction of the EoS: Section~\ref{sec:methods/eos/meta} briefly reviews the meta-model used for the low density EoS, and Section~\ref{sec:methods/eos/GP} presents our formalism for the high density Gaussian Process extensions. Section~\ref{sec:methods/mma} briefly reviews the EoS inference framework including details on hierarchical Bayesian analyses using astrophysical data. We present results in Section~\ref{sec:results} comparing the EoS distributions in pressure-density and characteristic macroscopic NS parameters with different methods of prior generation in Section~\ref{sec:results/mma/compare}. Finally, we revisit posterior EoS distribution informed by astrophysical observations in Section~\ref{sec:results/mma/current-constraints}.   

\section{Methodology}\label{sec:methods}

\subsection{Equation of state model} \label{sec:methods/eos}

Details about constructing our EoS prior using a semiparametric framework are presented here. At low densities, we employ the meta-model informed by nuclear experiment and theory. Beyond the stitching density, we generate GP extensions of the EoS trained with tabulated EoSs from the literature, specifically the set used in Ref.~\cite{Suleiman_2022}. We note that we do not use the piecewise polytrope fits of Ref.~\cite{Suleiman_2022}, but train our GP on the original nuclear theory EoSs collected in that work.

\subsubsection{Meta-model}\label{sec:methods/eos/meta}

Similarly to Ref.~\cite{Davis_2024}, we use the meta-model approach discussed in Ref.~\cite{Margueron_2018} to construct the low density EoS using the Crust Unified Tool for Equation of state Reconstruction (CUTER, see Zenodo repository in Ref.~\cite{Davis_2024}). In this non-relativistic approach, the energy per baryon of nucleonic matter $e$ at zero-temperature is expressed as a parabolic expansion in the isospin asymmetry parameter ${\delta = (n_n - n_p)/n}$
\begin{equation}
    e(n, \delta) = e_{\rm is}(n) + e_{\rm iv}(n) \delta^2 + t^*_{\rm FG}(n, \delta) \;,
\end{equation}
with $n$ the baryon density, $n_n$ and $n_p$ the neutron and proton density respectively, $e_{\rm is}$ the energy of isospin symmetric matter (also referred to as isoscalar energy), and $e_{\rm iv}$ the symmetry energy (also referred to as isovector energy). The parabolic expansion is corrected by a kinetic term which depends on several nuclear parameters: 
\begin{equation}
    t^*_{\rm FG}(n, \delta) = t^*_{\rm FG}(n, \delta;  n_{\rm sat}, m^*_{\rm sat}, \Delta m^*_{\rm sat}) \;,
\end{equation}
with $m^*_{\rm sat}$ the Landau effective mass in symmetric matter defined at nuclear saturation density $n_{\rm sat}$\footnote[1]{Note that the nuclear saturation density $n_{\rm sat}$ is not fixed, and should not be confused with the reference value for saturation density in lead, which is denoted $n_{\rm nuc}$. Here  $n_{\rm sat} \neq n_{\rm nuc}$.}, and the isosplit $\Delta m^*_{\rm sat}$ in neutron matter defined at $n_{\rm sat}$, see Eq.~(6) of Ref.~\cite{Margueron_2018}. Both the energy of symmetric matter and the symmetry energy are formulated as a fourth order truncated Taylor expansion around $n_{\rm sat}$ in which each order reveals additional parameters, so called nuclear empirical parameters, such that 
\begin{align}
    e_{\rm is}(n) &= e_{\rm is}(n; n_{\rm sat}, \{X_{\rm is}\}, m^*_{\rm sat}, \Delta m^*_{\rm sat}, b)\;, \\
    e_{\rm iv}(n) &= e_{\rm iv}(n; n_{\rm sat} , \{X_{\rm iv}\}, m^*_{\rm sat}, \Delta m^*_{\rm sat}, b)\;, \\ 
    \{X_{\rm is}\} &= \{E_{\rm sat}, K_{\rm sat}, Q_{\rm sat}, Z_{\rm sat} \} \;, \\
    \{X_{\rm iv}\} &= \{E_{\rm sym}, L_{\rm sym}, K_{\rm sym}, Q_{\rm sym}, Z_{\rm sym} \} \;.
\end{align}
We can identify here the isoscalar $\{X_{\rm is}\}$ and isovector $\{X_{\rm iv}\}$ nuclear empirical parameters defined at $n_{\rm sat}$: $E_{\rm sat}$ the symmetric matter (or isoscalar) energy, $K_{\rm sat}$ the isoscalar incompressibility, $Q_{\rm sat}$ the isoscalar skewness, $Z_{\rm sat}$ the isoscalar kurtosis, $E_{\rm sym}$ the symmetry energy, $L_{\rm sym}$ the slope of the symmetry energy, $K_{\rm sym}$ the isovector incompressibility, $Q_{\rm sym}$ the isovector skewness and $Z_{\rm sym}$ the isovector kurtosis. The parameter $b$ is introduced in Ref.~\cite{Margueron_2018} and ensures the correcting term to the potential energy quickly vanishes. For details on the explicit formulas for the kinetic term, the isoscalar and isovector energies, we refer to Eq.~(6-9) and Eq.~(16-24) of Ref.~\cite{Davis_2024} or to Eq.~(13-16) and Eq.~(22-31) of Ref.~\cite{Margueron_2018}.
\begin{table}[]
    \centering
    \begin{tabular}{|c||c|c|}
        \hline
        Parameter $X$ & $X_{\rm min}$ & $X_{\rm max}$ \\ \hline \hline 
        $n_{\rm sat}$ & 0.15 & 0.17\\ \hline 
        $E_{\rm sat}$ & -17.0& -15.0\\ \hline 
        $K_{\rm sat}$ & 190.0& 270.0\\ \hline 
        $Q_{\rm sat}$ & -1000& 1000 \\ \hline 
        $Z_{\rm sat}$ & -3000 & 3000\\ \hline 
        $E_{\rm sym}$ & 26.0& 38.0\\ \hline 
        $L_{\rm sym}$ & 10.0& 80.0\\ \hline 
        $K_{\rm sym}$ & -400 & 200 \\ \hline 
        $Q_{\rm sym}$ & -2000& 2000\\ \hline 
        $Z_{\rm sym}$ & -5000 & 5000\\ \hline
        $m^*/m$ & 0.6 & 0.8 \\ \hline
        $\Delta m^*/m$ & 0.0 & 0.2\\ \hline
        $b$ & 1.0 & 10.0\\ \hline
    \end{tabular}
    \caption{Meta-model nuclear parameters $X$ and the prior bounds from which they are sampled (following Ref.~\cite{Davis_2024}). The mean effective mass and effective isosplit are normalized to $m=\frac{m_n+m_p}{2}$, with $m_n$ and $m_p$ the neutron and proton mass respectively.}
    \label{tab:nuc_param}
\end{table}

We sample all the aforementioned nuclear parameters across a uniform distribution bounded by the experimentally informed values presented in Table~\ref{tab:nuc_param} (following Refs.~\cite{Dinh2021, Davis_2024}). We apply nuclear theory constraints on the energy per nucleon of pure neutron matter using $\chi$EFT calculations presented in Ref.~\cite{2016PhRvC..93e4314D} in the interval of baryon density ${n\in[0.02\;{\rm fm}^{-3}, n_{\rm nuc}]}$. After adding the leptonic contribution, we compute a set of $\beta$-equilibrated EoSs up to the fixed reference value for saturation density $n_{\rm nuc} = 0.16$\;fm$^{-3}$ according to the method described in Ref.~\cite{Davis_2024}. The crust inhomogeneities are treated within the Cold Compressible Liquid Drop Model as described in Ref.~\cite{Carreau2019}, and the surface parameters are fitted to the atomic mass table AME2020 \cite{ame2020} with the method presented in Ref.~\cite{Davis_2024}. The core crust transition is calculated consistently using the CLDM and meta-model approach with a minimum energy compared to homogeneous and inhomogeneous matter.

We use a fourth order truncated Taylor expansion in the meta-model. As discussed in Ref.~\cite{Margueron_2018} (Section III.E)  the meta-model may fail to reproduce realistic EoSs at high density where the higher order terms become important. Additionally, by construction, the meta-model describes only nucleonic matter, but it has been hypothesized that ``exotic" degrees of freedom such as hyperons (baryons with a strange quark) or deconfined quarks could exist in the core of NSs. For these reasons, we transition into a different representation of the EoS at densities $n \geq n_{\rm tr}=n_{\rm nuc}$. We discuss the choice of this transition density in more detail in Appendix \ref{sec:appendix/prior-construction}. While piecewise polytropes were used with the meta-model in Ref.~\cite{Davis_2024} to model the EoS beyond $n_{\rm tr}$, here we use Gaussian Process extensions for an agnostic high density EoS. 

\subsubsection{Gaussian Process formalism} \label{sec:methods/eos/GP}

There are many different approaches for the high density EoS in the literature (see Sec. 5 of Ref.~\cite{Oertel_2017}) leading to potential EoSs with a variety of theoretically or phenomenologically motivated functional forms. Here, we turn to a nonparametric approach for constructing EoSs at densities past the stitching density $\rho_{\rm tr}$. We use Gaussian Processes (GPs) which are stochastic processes with the ability to encode the uncertainties in a function. GPs demonstrate increased model freedom over parameterized EoS \cite{Legred_2022}, in that they do not assume a particular functional form of the EoS \textit{a priori}, and allow for non-zero probabilities for any causal and thermodynamically stable EoSs. 

Our high density EoS is constructed to describe the pressure $p$ as a function of $\rho$, the baryonic rest-mass density, using an average baryon mass such that our transition density $n_{\rm tr}=n_{\rm nuc}$ corresponds to 
$\rho_{\rm nuc} = 2.65\times10^{14}$ g/cm$^3$ following \cite{Tsang_2024, Lynch_2022}.

Within GPs, a real-valued, continuous function $f$ is taken as an element of an infinite-dimensional vector space, such that the GP gives a joint probability density for the function values $f_{i} \coloneqq f(x_{i})$ over a domain $\vec{x} = \{x_{i}\}$, where correlations in function values are modeled as a multivariate Gaussian distribution $\mathcal{N}$. The distribution is described by two moments: its mean function $\vec{\mu}$ and covariance matrix $\Sigma$. The GP then takes the form of
\begin{equation}
    f(\vec{x}) \sim \mathcal{N}(\vec{\mu}, \Sigma_{ij}),
\end{equation}
where elements of the covariance matrix are computed using covariance functions (kernels) and the mean $\vec{\mu}$ is the expectation value of a function $f$ such that,
\begin{align}
    \vec{\mu} = \langle f \rangle = \mathbb{E}[f] \;, \\
    \Sigma_{ij} = K(x_{i},x_{j}) \;.
\end{align}

To generate EoSs using the GP formalism, we follow the works of Refs.~\cite{Landry_2019,EssickLandryHolz_2020} in that our GP is built on an auxiliary variable introduced in Ref.~\cite{Lindblom_2010}, namely $\phi$ given by,
\begin{equation}
    \phi\left(\log \rho \right) \equiv \ln\left(c^{2}\frac{d\varepsilon}{dp} - 1\right)
\end{equation}

where $\varepsilon$ is the energy density. Although it is possible to immediately sample from the pressure-density plane directly, realizations from a GP enacted onto the EoS function space may yield EoSs exhibiting acausality and thermodynamic instability, so it is in our interest to sample in the $\phi$ - $\rho$ plane.  For $\phi \in \mathbb{R}$, by construction, the sound speed of a corresponding EoS will remain causal $c_{s} = \sqrt{dp/d\varepsilon} \leq c$ and thermodynamic stability $dp/d\varepsilon \geq 0$ is enforced.  

We use kernels comprised of hyperparameters to create a covariance matrix and modify the covariances between function values, altering model attributes such as correlation length scales and ultimately obtain prior support across a broad range of pressures for samples drawn from the GP. The specific hyperparameter choices made in constructing our EoS prior, and their implicated effects are discussed in Appendix \ref{sec:appendix/hyperparameters}. Common kernels used in GP regression methods include the squared-exponential kernel \cite{Rasmussen_2005}, which generates smooth functions sampled by the GP with variations on a chosen length scale.

In this work, we implement the rational quadratic kernel $K_{\rm rq}$~\cite{duvenaud2014automatic},
\begin{equation}
    K_{\rm rq}(x_{i},x_{j}) = \gamma^{2}\left(1 + \frac{|x_{i}-x_{j}|^{2}}{2\alpha \ell^{2}} \right)^{-\alpha}
\end{equation}
which can be interpreted as an infinite sum of squared-exponential covariance functions with dissimilar length scales. Here, $\gamma$ is the overall covariance strength, $\alpha$ is a measure of scale-mixture, and $\ell$ is the characteristic length scale for correlations. By implementing the rational quadratic kernel, we effectively allow the functions generated by the GP to vary across multiple length scales. 

Additionally, we incorporate a model uncertainty kernel denoted $K_{\rm ts}$ into our covariance matrix, computed from nuclear theory input training models,
\begin{equation}
    K_{\rm ts}(x_{i}, x_{j}) = \frac{1}{N^{(\nu)}}\left((\phi^{(\nu)}_{i} - \langle \phi^{(\nu)}_{i}\rangle)(\phi^{(\nu)}_{j} - \langle \phi^{(\nu)}_{j}\rangle ) \right)
\end{equation}

where quantities with $(\nu)$ indicate relation to nuclear theory models. Here $N^{(\nu)}$ represents the total number of training input models, and $\phi^{(\nu)}_{i}$ denotes a nuclear theory training model's $\phi$ value at a given density point. 
Included in the nuclear training data are EoSs with nucleons, hyperons and phase transition to deconfined quarks in the core, allowing our GP to emulate a variety of features.
The specific selection of nuclear theory models used for GP training is described in Sec 2. of Ref.~\cite{Suleiman_2022}.

In this work, the GP region acts as an extension to meta-model EoSs at low density, so we condition our GP on the existence of the meta-model EoS at $\rho_{\rm tr}$. For a GP to sample from a conditional joint probability distribution, the covariance matrix $\Sigma$ can be decomposed into 4 sub-matrices,
\begin{equation}
\Sigma = 
    \begin{bmatrix}
    K & K^{*} \\
    (K^{*})^{T} & K^{**} \\
    \end{bmatrix}
\end{equation}
where $K$ holds the covariances for the testing points (where the GP predicts values at before conditioning), $K^{*}$ contains the covariances between testing and conditioning points, and $K^{**}$ is the covariance matrix for the conditioning point(s)\footnote[2]{Quantities with asterisks indicate domain points with which we condition the Gaussian Process at.}.
The covariance matrix $\Sigma$ is of dimensions $N\times N$; we denote $q$ the number of points at which we condition the GP on. We condition the GP on a given meta-model only at $\rho_{\rm tr}$, therefore $q = 1$. $K^{**}$ is then a matrix of dimensions $q \times q$, $K^{*}$ of dimension $(N-q) \times q$, and $K$ of dimension $(N-q) \times (N-q)$. Modifying the original GP to sample from the conditional distribution (see Appendix A from Ref.~\cite{Rasmussen_2005} for a derivation), the resulting GP assumes the form,

\begin{equation}
\begin{split}
    \phi_{i}(\log\rho) & \sim \mathcal{N} \bigg(\langle \phi_{i} \rangle + 
K_{ik}^{*}(K^{**})^{-1}_{kj}\left(\phi^{*}_{j} - \langle\phi^{*}_{j}\rangle \right), \\
    & K_{ij} - K_{im}(K^{**})^{-1}_{mn}K^{*}_{nj} \bigg) \;.
\end{split}
\end{equation}

Once we have our set of $\phi_{i}(\log \rho)$, we recover the GP EoS extension's energy densities and pressures using the first law of thermodynamics in the zero temperature limit,
\begin{equation}
    d\varepsilon = \frac{p + \varepsilon}{\rho}d\rho \;.
\end{equation}

\subsection{Astrophysical observations}\label{sec:methods/mma}

In the assumption of non-spinning NSs in general relativity, macroscopic properties of neutron stars such as their masses, radii, and tidal deformability, are uniquely determined by the EoS, the central density, Tolman-Oppenheimer-Volkoff (TOV) equations \cite{Oppenheimer-Volkoff_1939,Tolman_1939} and the Love number equations \cite{Hinderer_2010, Damour_2009}. We can therefore constrain the NS EoS using observations of macroscopic parameters. Observations most predominantly used for constraining the EoS by astrophysical means are radio measurements of mass for massive pulsars (PSRs), gravitational-waves (GWs) from binary neutron star mergers, and X-ray pulse profile modeling of millisecond pulsars (X-ray). 

In this work, we use a hierarchical Bayesian framework to impose the astrophysical constraints from each class of observation on the EoS set; our approach is similar to the framework discussed in the works of Refs.~\cite{Landry_2020, Legred_2021}. The probability likelihood for a given EoS is comprised of the marginal likelihoods obtained from each set of observations. In this regard, the posterior probability for a given EoS is given by, 

\begin{equation}
    \mathcal{P}(\epsilon|d) = \frac{\mathcal{P}(d|\epsilon)\mathcal{P}(\epsilon)}{\mathcal{P}(d)}
\end{equation}
where $\mathcal{P}(\epsilon)$ is the prior on the EoS, $\mathcal{P}(d|\epsilon)$ the likelihood of some observational data $d$ given an EoS $\epsilon$, $\mathcal{P}(d)$ the evidence, and $\mathcal{P}(\epsilon|d)$ is the posterior. The marginalized likelihood from all observations can be obtained from the relation,

\begin{equation}
    \mathcal{P}(d|\epsilon) = \mathcal{L}(\vec{\theta}) = \prod_{i} \mathcal{L}_{i}(\vec{\theta})
\end{equation}
with $\mathcal{L}_{i}$ indicating the probability likelihood from each observation class consisting of PSR, GW, or X-ray. Using \lwp ~\cite{LWP,EssickLandryHolz_2020,Landry_2019,Landry_2020,Legred_2021,Legred_2022}, an inference code for computing marginal likelihoods of EoS based on gravitational wave data, we constrain our EoS distribution accordingly. 

\subsubsection{Radio}\label{sec:methods/mma/radio}

Observed masses of heavy pulsars in binaries limit the lower bound for the maximum mass of NSs, acting as a threshold for each EoS to minimally support.
When constraining EoSs with massive pulsar measurements, a standard technique is to perform rejection sampling for any EoS that does not support the observed mass. In our framework, we represent the probability likelihood of the mass measurement as a Gaussian with the tails set to the uncertainty values reported by the observation. As in the case of PSR J0348$+$0432, with a mass of $m = 2.01 \pm 0.04 M_{\odot}$ at 1-$\sigma$, each EoS of the posterior is compatible with a maximum mass above this mass threshold. In our study, we use pulsar observations PSR J0348$+$0432 \cite{Antoniadis_2013}, PSR J0740$+$6620 \cite{Fonseca:2021wxt}, and PSR J1614$-$2230 \cite{Arzoumanian_2018}.

\subsubsection{Gravitational-waves}\label{sec:methods/mma/gw}

During the late stages of a BNS merger inspiral, the internal structure of the NSs prescribed by their EoS begin to affect both the orbital phase of the binary and amplitude of the emitted gravitational waves \cite{Chatziioannou_2020}. Most dominantly, the effects on the phase accelerates the coalescence due to the external tidal field exerted onto one star from the other. As a measure of the deforming effect, the parameterized dimensionless tidal deformability $\Lambda$ is given by,
\begin{equation}
    \Lambda = \frac{2}{3}k_{2}\left(\frac{R}{m}\right)^{5}
\end{equation}
where $k_{2}$ is the Love number correspondent to the $l = 2$ leading order perturbation \cite{Hinderer_2010}.

When detecting gravitational wave signals, the most accurately measured quantity is the gravitational wave phase which explicitly depends on $\Lambda_{1,2}$ to leading order. Both $k_{2}$ and $R$ are EoS dependent, and it is therefore possible to constrain the NS EoS directly from gravitational wave inference of $\Lambda$ values.
To date, the two BNS mergers that have been detected by advanced LIGO and Virgo are GW170817 \cite{Abbott_2017, Abbott_opendata_2021, Abbott_2019_GW17properties} and GW190425 \cite{Abbott_2020}. Although our analysis includes GW190425, tidal constraints from this event are minimal and show close to no effect in the context of the hierarchical inference when in comparison to the inclusion of tidal information obtained from GW170817. 

\subsubsection{X-ray}\label{sec:methods/mma/xray}

Millisecond pulsars can emit pulsed soft x-ray signals from hotspots on their surfaces. Through pulse profile modeling, which describes both the thermal x-ray emission and its transmission through the spacetime surrounding the neutron star, the mass and radius of the object can be inferred \cite{Bogdanov:2021yip}.

The Neutron Star Interior Composition ExploreR (NICER) has collected observational data for several nearby pulsars. Observations of the pulsar PSR J0030$+$0451  with a mass $\sim 1.3 - 1.7 M_\odot$, have led to radius estimates from 11.7 to 14.4 km \cite{Miller_2019,Riley_2019,Vinciguerra_2024}. The massive pulsar PSR J0740$+$6620 has a prior mass measurement of $2.08 \pm 0.07 M_\odot$ from pulsar-timing in radio \cite{Fonseca:2021wxt}, and the incorporation of pulse profile modeling led to large radius estimates of 12.4 to 13.7 km \cite{miller_2021radius,riley_2021nicer,Salmi_2024}. Additionally, X-ray pulse profile and radio pulsar timing observations of PSR J0437-4715 have led to a mass measurement of $1.418\pm0.037~M_\odot$ and radius of $11.36^{+0.95}_{-0.63}$~km \cite{Choudhury_2024} from pulsar timing in radio \cite{Reardon_2024}. The most recent observation of millisecond pulsar PSR J0614-3329 \cite{Mauviard_2025} in X-ray and radio has led to an inferred mass of $1.44^{+0.06}_{-0.07} M_{\odot}$ and radius estimates of $10.29^{+1.01}_{-0.86}$\;km. Our analysis uses x-ray observations of J0030$+$0451~\cite{Miller_2019,Vinciguerra_2024}, J0740~\cite{Dittmann_2024,Salmi_2024}, J0437-4715~\cite{Choudhury_2024}, and J0614-3329~\cite{Mauviard_2025}.

\subsubsection{The population of compact objects}
Generically, the population of NSs must  be inferred simultaneously with the EoS in order to avoid biases~\cite{Wysocki_2020}. Nonetheless, if the number of observations is small, this bias will also be small; therefore, in this work we fix the population of compact objects. Since we do not know for certain that any particular compact object in a gravitational-wave binary is a NS, we assume that the components of GW170817 and GW190425 are NSs if their masses are less than the TOV maximum mass for a given EoS. We assume therefore that the population of merging compact objects is uniform on $m_1, m_2 \in (1.0, 3.0)\, M_{\odot}$, where the range contains the bulk of the posterior mass distribution for the observed events. 
On the other hand, pulsars are known to be NSs, therefore when we sample the mass, we require that the TOV mass limit of a given EoS exceed the sampled mass in order to assign it nonzero likelihood.
We assume the galactic NS population is also uniform, although because we require the population maximum mass to be less than the TOV maximum mass,  the population maximum mass will generically depend on the EoS. We assume that pulsars like J0740$+$6620 and J0348$+$0432 are formed up to the TOV maximum mass, since they are identified as candidates for the maximum NS mass, but that J0437-4715, J0614-3329, and J0030$+$0451 come from a uniform population with mass inference that is not restricted by the TOV maximum mass. The primary effect is that only J0740$+$6620 and J0348$+$0432 include an Occam penalty associated with the increased prior volume since the density of the prior for these sources goes like $\pi(m) \propto 1/(M_{\rm TOV} - M_{\min})$, where we choose $M_{\min} = 1.0\, M_{\odot}$; see Refs.~\cite{Landry_2020, Legred_2021, Golomb:2024lds} for additional details of this method.

\begin{figure}[t!]
    \centering
    \includegraphics[width=\columnwidth]{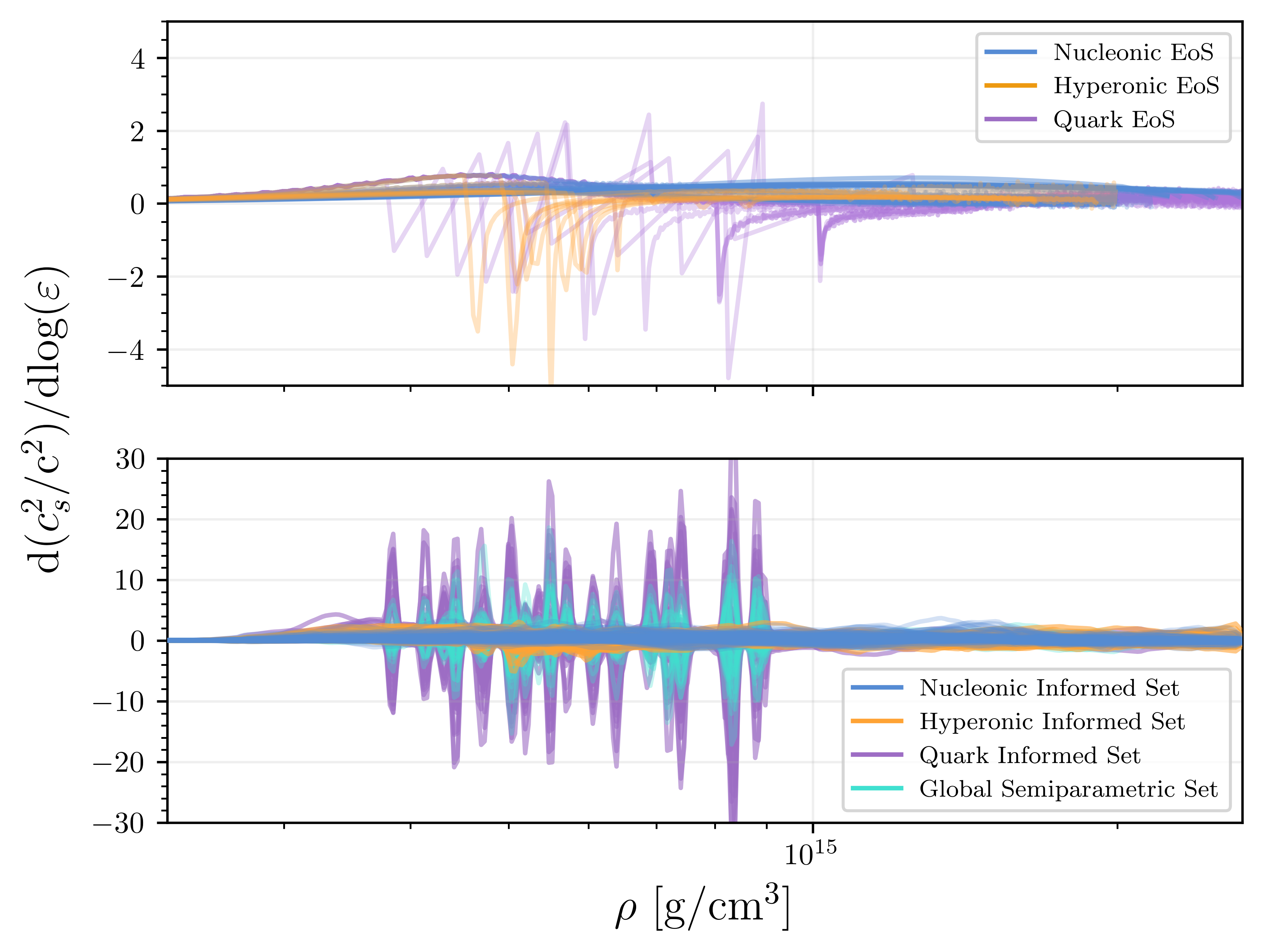}
    \caption{Semiparametric models trained on different types of core compositions and their rate of change of their sound speeds, with respect to logarithmic energy density. All training EoS are shown in the top panel. Specific composition informed draws, and draws from the overarching distribution are shown in the bottom panel.}
    \label{fig:changes-in-sound-speed}
\end{figure}

\section{Results} \label{sec:results}

\subsection{Semiparametric extensions}\label{sec:results/eos}
\subsubsection{Impact of training data}\label{sec:results/training-data}

To examine the impact of the training data on the GP extensions, we generate separate sets of semiparametric models with their GP extensions trained on subsets of the nuclear training data EoSs according to their respective core-composition groups: nucleonic, hyperonic and quarkyonic. By doing so, we isolate and check emulated features within our EoS prior compared to behaviors in the respective training EoS subset, keeping in mind that the total semiparametric EoS prior will have hybridized features among all three trained groups.

Phase transitions in an EoS can appear as drops in $c_{s}^{2}$ \cite{Essick_2023, Mroczek_2024, Brandes_2024}, which would explicitly show behavior in the one-to-one mapping to $\phi$, and in turn alter the covariance of the subsequent GP. In the GP-generated density regions, we consider how well our semiparametric models can reproduce features shown by EoS that include phase transitions, specifically changes in the sound speed with respect to energy density. As can be seen in Fig.~\ref{fig:changes-in-sound-speed} (top panel), training EoSs with phase transitions to quark matter show features in the speed of sound derivative that are absent in nucleonic core EoSs. These features occur at the densities of phase transitions in the training EoSs.

We observe from Fig.~\ref{fig:changes-in-sound-speed} (bottom panel) that the semiparametric draws generated from the GP trained on tabulated nuclear theory EoSs with phase transitions also present sharp jumps and drops in the sound speed derivative. The impact on the overall pressure-density relationship from incorporating quark core EoSs in the training data is a broadening of the range of pressure-density space covered by our EoS set.  
These sharp sound speed derivative features are not present for GPs trained on nucleonic tabulated EoS, and are very limited for the training on hyperonic core tabulated EoSs. The global semiparametric set trained on all tabulated EoSs contains some draws with sharp features in the sound speed derivative. 
Note that the difference in scaling in the y-axis between the training data and the different informed processes; our GP generates some extreme draws with a significantly larger variability as it samples from the EoS distribution.

Ultimately, we include all EoS categories in the training data for our Gaussian process. We do this because we cannot necessarily verify that individual choices of training data will create processes which emulate the training EoSs; e.g. that a quark matter-conditioned GP will faithfully emulate quark-matter EoSs. Fundamentally, the training EoSs do not follow an underlying statistical distribution, and therefore a GP conditioned on them cannot be said to emulate any particular distribution. Nonetheless, we suggest that the overall distribution, which is conditioned on hadronic, hyperonic, and quark EoS, can act as an emulator for the global space of all EoSs including those with phase transitions.

\begin{figure}[t!]
    \centering
    \includegraphics[width=\columnwidth]{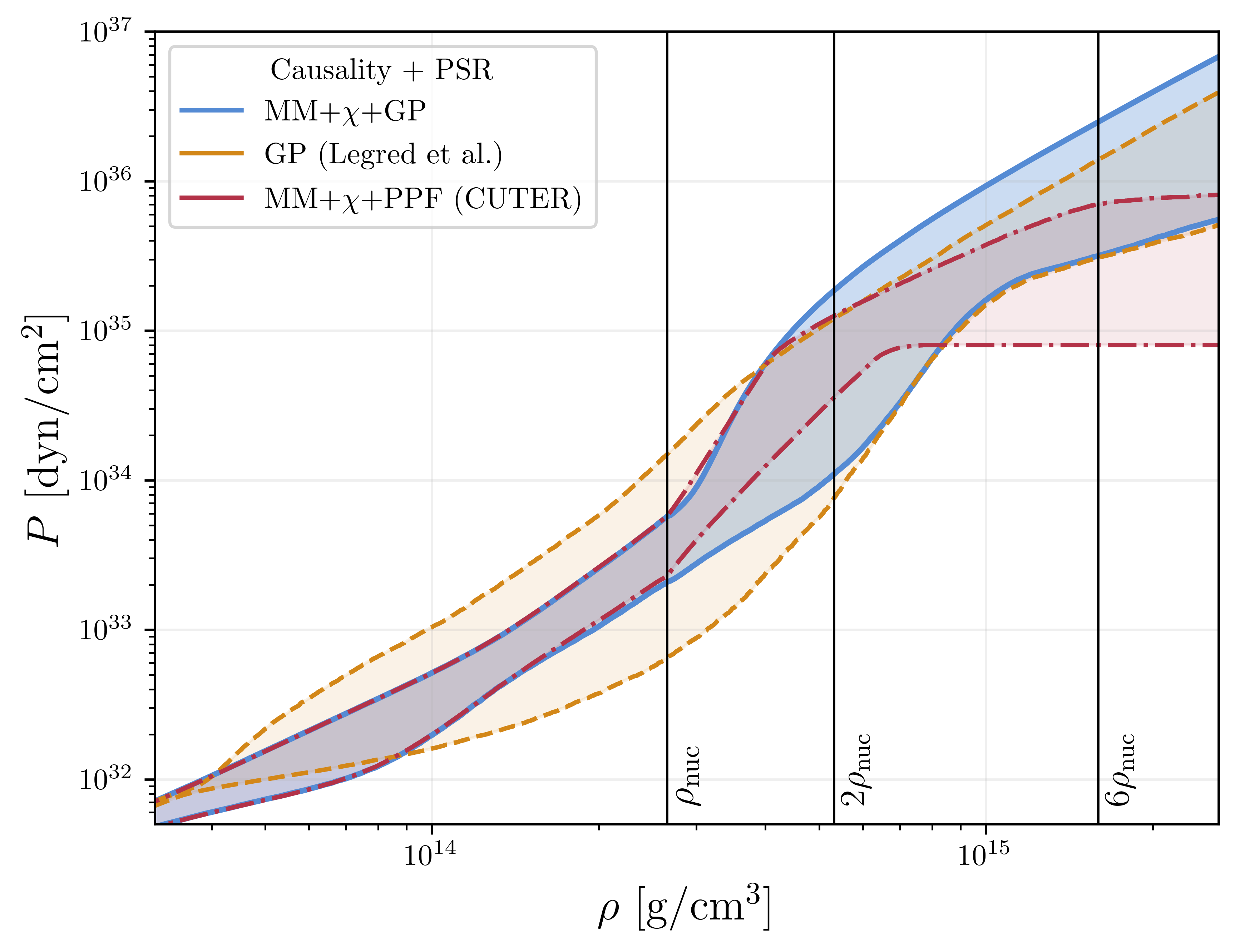}
    \caption{Pressure-density relation of multiple EoS distributions at 90\% C.L limits adhering to causality and measurements of heavy pulsars. The semiparametric EoS distribution in blue and a set of meta-model EoS with piecewise polytropic extensions in dark red, both stitched at $\rho_{\rm tr}$, are shown. Additionally, we display the Legred et al. model-agnostic nonparametric EoS in gold.}
    \label{fig:low-density-comparison}
\end{figure}

\subsubsection{Impact of Modeling}
\label{sec:results/mma/compare}

When comparing the flexibility of EoS models, we found it useful to impose certain astrophysical constraints in order to understand the distribution of "astrophysically plausible" EoSs produced from each model. Following arguments in Ref.~\cite{Legred_2022}, we first condition our EoS distributions on mass measurements of heavy pulsars. Our comparison between EoS generation models then enforces two clear and unambiguous constraints which bracket the understanding of high density EoS: causality, which sets overall upper limits on pressures, and observed pulsar masses, which sets overall lower limits on pressures. The breadth of allowed EoS after these two restrictions have been imposed is a clearer measure of the flexibility of the modeling framework than a full prior range, which can typically be arbitrarily extended by modifying the range of model (hyper)parameters. 
Our comparison shows that both incorporating the nuclear physics constraints below nuclear saturation density and changing the method of constructing the EoS at high density have significant impacts on the range of high density pressure distributions.

In Fig.~\ref{fig:low-density-comparison}, we show the EoS distribution for our semiparametric construction (MM+$\chi$+GP) with a $\chi$EFT constrained meta-model at low density and GP high density extensions. We compare it to (i) the set of semi-agnostic models (generated with CUTER) presented in Ref.~\cite{Davis_2024} (MM+$\chi$+PPF) based on the same low density treatment but with piecewise polytropic high density extensions using 5 density segments and, (ii) the set of nonparametric EoSs presented in Ref.~\cite{Legred_2021} (GP Legred et al.). 

Below nuclear saturation density, our semiparametric framework reproduces the pressure contours of the meta-model, using information from nuclear experiment and $\chi$EFT. In contrast, the  nonparametric EoS covers a wider range at $\rho_{\rm nuc}$: the upper and lower bounds on the pressure of the fully nonparametric EoS distribution yields $\Delta P = 1.43\times10^{34}$ dyn/cm$^2$, whereas the limits on pressure from the meta-model produce $\Delta P = 3.63\times10^{33}$ dyn/cm$^{2}$, all at the 90\% credible level (C.L). 

Between 1-2$\rho_{\rm nuc}$, the GP extension in the semiparametric EoS is considerably restricted by the meta-model connection, as thermodynamic consistency limits the rate of changes in pressure. Noticeably, both the piecewise polytropes and semiparametric GP extensions have general agreement on the upper bound of pressure up to $1.5\rho_{\rm nuc}$; this suggests these EoS in each case explore the possible range up to the causal limit. However, the lower bound of the piecewise polytropic extensions is notably more restrictive compared to nonparametric and semiparametric EoSs. Since these piecewise polytropes have coarse-grained changes in adiabatic index and are not causal by construction, such that causality must be imposed \textit{a posteriori}, we interpret the preference for higher pressures in this density range as driven by the inherent inflexibility of the piecewise polytrope extensions in reaching the heavy pulsar constraints without breaking causality at high densities. 
From 2$\rho_{\rm nuc}$ to the end of the density range presented in Fig.~\ref{fig:low-density-comparison}, the lower bound of the piecewise polytrope extension approaches a constant, as would be set by thermodynamic consistency requiring nondecreasing pressure with density. We interpret this as the requirement to support massive pulsars has been satisfied by the preference for higher pressure at 2$\rho_{\rm nuc}$, so thermodynamic consistency alone sets the lower bound in this region.

Of the two GP models used at high density, our semiparametric EoS set allows a higher upper bound on pressure above $1.5\rho_{\rm nuc}$ than the nonparametric distribution, which we attribute to tuning our GP specifically for a broad prior range of pressures compatible with meta-model constraints as discussed in Appendix~\ref{sec:appendix/hyperparameters}. We targeted flexibility in our GP by generating EoS draws that vary on multiple scales through a rational quadratic kernel, and chose hyperparameters for the kernel that lead to the widest range of pressures. The nonparametric GP we compare to is also built in $\phi(\log P)$ but with different kernel implementations, and only loosely trains on nuclear theory EoS, incorporating a white-noise kernel that generates a prior distribution extending to very low pressures at nuclear saturation density and therefore its 90\% contours also tend toward lower pressures. In the interval of $2\rho_{\rm nuc}$ and $6\rho_{\rm nuc}$, the fully nonparametric EoS distribution and semiparametric EoS distributions agree in their lower bound. This suggest both semiparametric and nonparametric EoS are fully exploring the lowest-pressure thermodynamically consistent EoS that support massive pulsars in this density range. 

\begin{figure}[t!]
    \centering
    \includegraphics[width=\columnwidth]{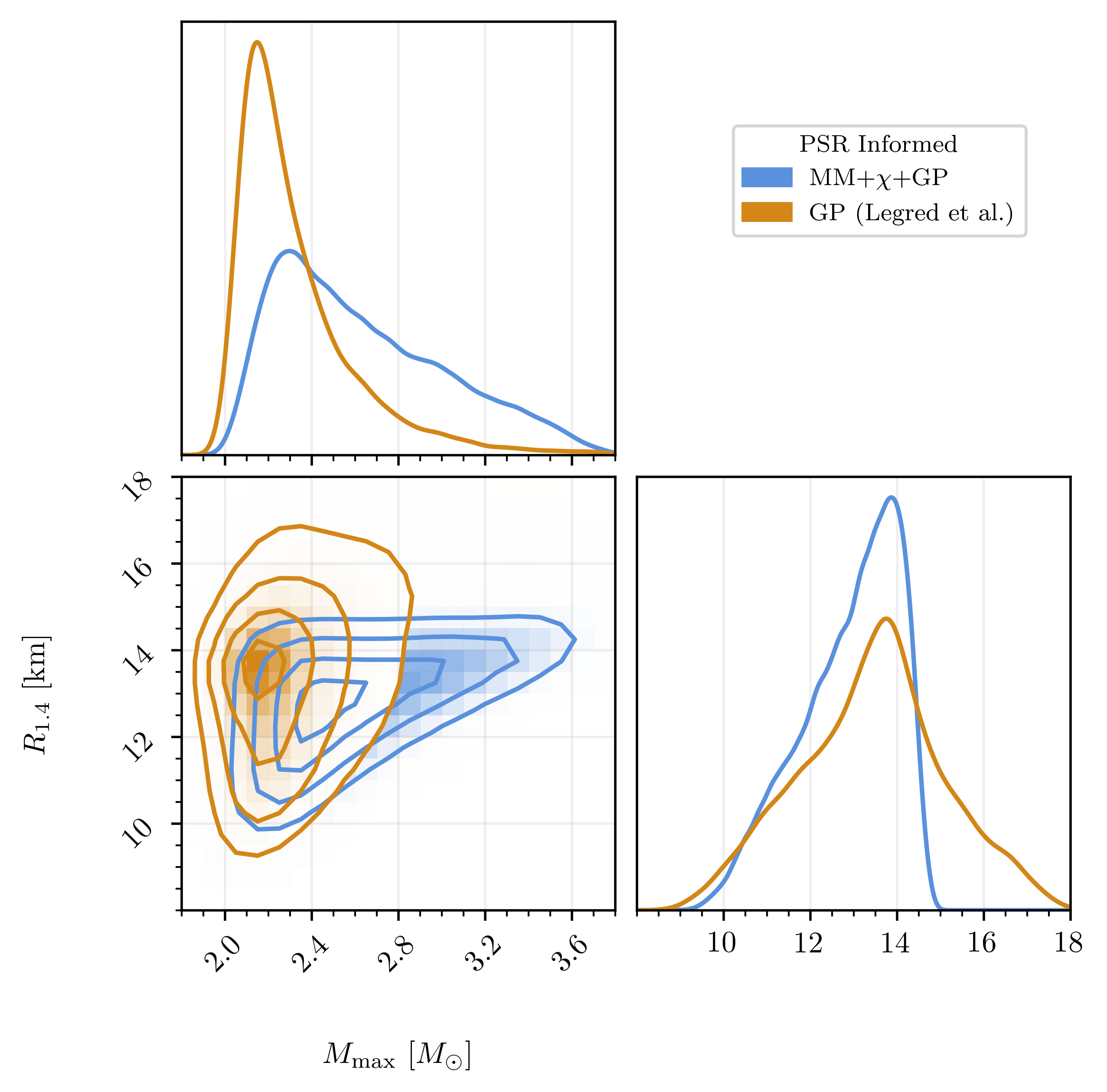}
    \caption{Comparison of the 90\% C.L of $R_{1.4}$ and $M_{\rm max}$ as supported by both the semiparametric EoS model (blue), and the fully nonparametric EoS model (gold). Both EoS distributions are constrained with heavy pulsar mass measurements from observations PSR J0740$+$6620 and PSR J0348$+$0432.}
    \label{fig:semiparam-gp-correlation}
\end{figure}

\begin{figure*}[t!]
    \centering
    \begin{subfigure}[h]{0.49\textwidth}
        \centering
        \includegraphics[width=\linewidth]{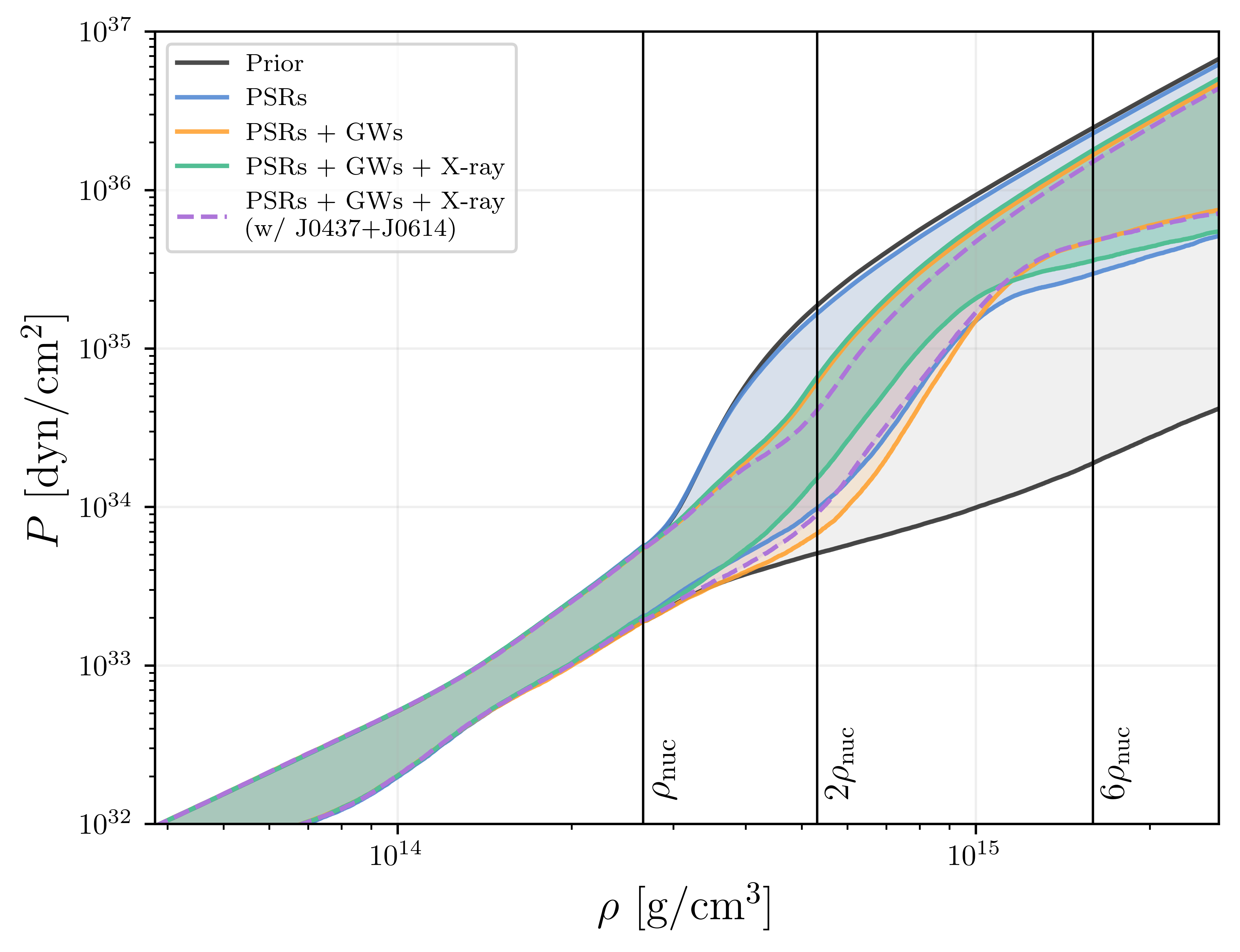}
        \phantomcaption{}
        \label{fig:all-astro-press-dens}
    \end{subfigure}
    \begin{subfigure}[h]{0.49\textwidth}
        \centering
        \includegraphics[width=\linewidth]{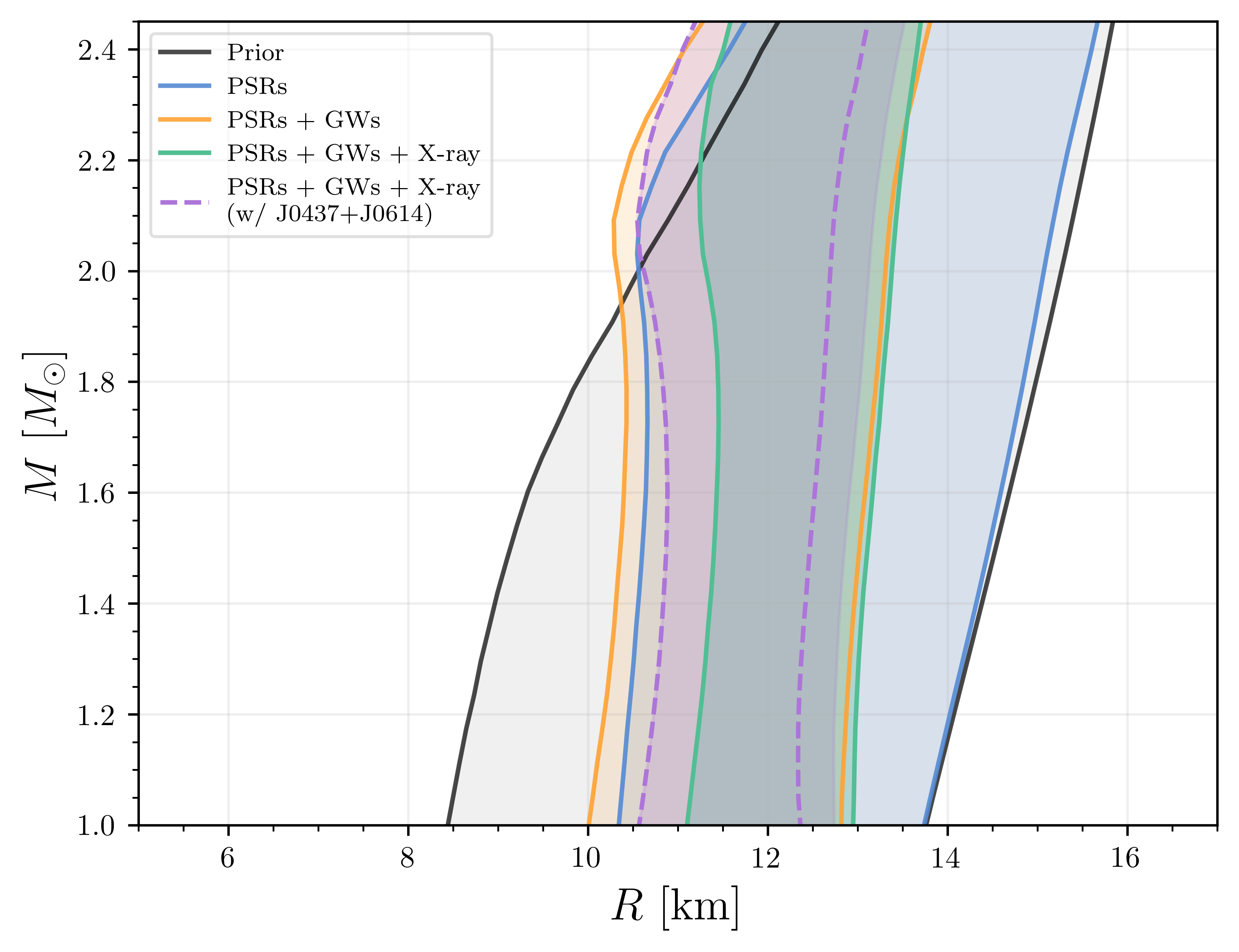}
        \phantomcaption{}
        \label{fig:all-astro-mass-radius}
    \end{subfigure}
    \caption{Stacking constraints from astrophysical observations of massive pulsars (blue), gravitational waves (orange), and X-ray emissions from millisecond pulsars (green and purple), with bounds on pressure as a function of density (left) and on radius as a function of mass (right). 1-D symmetric bounds of the semiparametric EoS model are shown at 90\% credible levels for each density and each pressure. Green lines represent our semiparametric EoS posterior with the NICER constraints of J0030+0451 and J0740+6620 \cite{Miller_2019,Dittmann_2024}. Purple contour lines denote our semiparametric EoS posterior with NICER measurements including J0437-4715 and J0614-3329 \cite{Vinciguerra_2024,Salmi_2024,Choudhury_2024,Mauviard_2025}. }
    \label{fig:all-astro-constraints-total}
\end{figure*}

In Fig.~\ref{fig:semiparam-gp-correlation}, we display a comparison between astrophysically relevant macroscopic quantities derived from our semiparametric EoSs and those generated by the nonparametric EoSs: the radii at $1.4 M_{\odot}$ denoted $R_{1.4}$, and the maximum neutron star mass supported by a given EoS, M$_{\rm max}$. These results are conditioned on the same observations and use the same EoS distributions as seen in Fig.~\ref{fig:low-density-comparison}. We see that the use of the nuclear-physics informed meta-model at sub-saturation densities induces a stringent limit on $R_{1.4}$ at $\sim 14$\; km, whereas the Legred et al. samples supports EoS with radii up to $\sim 17$\; km. This difference arises from the reduced upper limits on pressures below approximately $1.5\rho_{\rm nuc}$ in the meta-model part of our semiparametric EoS, as required by experimental and $\chi$EFT constraints at $\rho_{\rm nuc}$ and causality. We note that radius reflects both the core and crust of the candidate $1.4 M_{\odot}$ stars, and is therefore more sensitive to nuclear-density physics.

In contrast to the increased restrictions on radius, the increased upper limit on pressures explored by causal EoS in our semiparametric model at $2$ and $6\rho_{\rm nuc}$ results in a shifted and extended prior range for $M_{\rm max}$ that prefers higher values, as seen in the top panel of Fig.~\ref{fig:semiparam-gp-correlation}. The GP generates causal EoS candidates that support larger masses than is typically seen in nuclear theory models, e.g. in the covariant density functional approach of \cite{Li:2025oxi}. Although our range of maximum masses extends past $3.0 M_{\odot}$, it is consistent with previously published theoretical bounds such as those in Ref.~\cite{Kalogera_1996}, as the choice of crust EoS and stitching density affects the inferred maximum NS mass in that work. 

\subsection{Astrophysical constraints}\label{sec:results/mma}

\subsubsection{Results for the semiparametric EoS}
\label{sec:results/mma/current-constraints}
Cumulative constraints on the pressure-density and mass-radii relations are obtained with the Bayesian hierarchical analysis framework as briefly discussed in Sec.~\ref{sec:methods/mma}. In Fig.~\ref{fig:all-astro-press-dens}, we obtain posteriors on the EoS in the pressure-density plane. As discussed in Section \ref{sec:methods/mma/radio}, measurements of pulsar masses require higher pressures, thus resulting in a posterior distribution disfavoring lower pressures in the high-density EoS prior.
When we fold in additional constraints from gravitational wave measurements GW170817 and GW190425, we soften the EoS posterior between $1.5\rho_{\rm nuc}$ and $\sim4\rho_{\rm nuc}$ and disfavor higher pressures. While this pattern is still broadly consistent with early parametric results \cite{LIGOScientific:2018cki}, our more flexible model has broadened the pressure-density ranges compatible with the observation, as seen with the GP results of Ref.~\cite{Landry_2020}.

We consider two options for weighting the EoS likelihood using the NICER X-ray constraints, to demonstrate the impact of recent observations using NICER data. 
We first consider the impact of the first two pulsars observed by NICER, J0030$+$0451 as analyzed in Ref.~\cite{Miller_2019}, and J0740$+$6620 as analyzed with additional data in Ref.~\cite{Dittmann_2024}. The J0740$+$6620 X-ray constraint includes and replaces the radio pulsar mass constraint for the same source. As these observations support moderately large radii, they increase the lower bound on pressures above nuclear saturation density.

We then consider the alternative mass-radii measurements of PSR J0030$+$0451 in Ref.~\cite{Vinciguerra_2024} (\texttt{ST+PST} model) and PSR J0740$+$6620 from Ref.~\cite{Salmi_2024}, and then add in the more recent PSR J0437-4715 from Ref.~\cite{Choudhury_2024}, and PSR J0614-3329 as discussed in Ref.~\cite{Mauviard_2025}. 
The inference of J0437-4715 and J0614-3329 \cite{Choudhury_2024, Mauviard_2025} yield roughly similar inferred masses and radii to the gravitational-wave signal GW170817 \cite{LIGOScientific:2018cki}, and return the EoS distribution to the center of the PSR + GW posterior region. Including these events leads to lower pressures between $1.5\rho_{\rm nuc}$ and $4\rho_{\rm nuc}$, and requiring higher pressures at densities above $4\rho_{\rm nuc}$ to continue to support massive stars. 

Taking into account all discussed astrophysical constraints, we report a median pressure at $2\rho_{\rm nuc}$ of $P = 1.98^{+2.13}_{-1.08}\times10^{34}$ dyn/cm$^{2}$. 
 
\begin{figure}[t!]
    \includegraphics[width=\columnwidth]{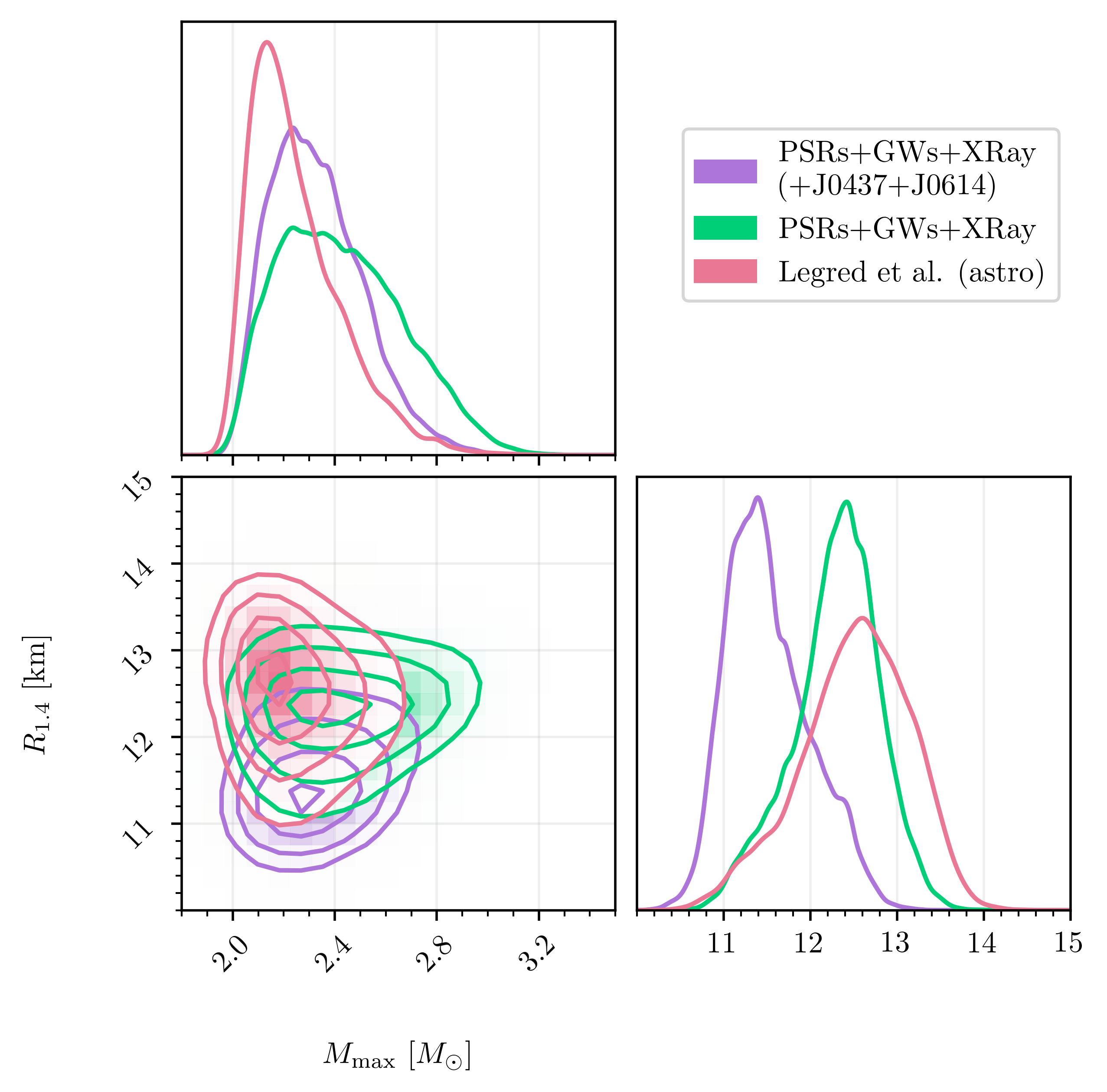}
    \caption{Posterior distributions of the semiparametric EoS compared with the astrophysically constrained nonparametric EoS posterior of Legred et al., which includes earlier NICER results for J0030+0451 and J0740+6620 only. Pink lines correspond to that previous nonparametric posterior. Green lines represent our semiparametric EoS posterior with the NICER constraints of J0030+0451 and J0740+6620 \cite{Miller_2019,Dittmann_2024}. Purple contour lines denote our semiparametric EoS posterior with all public NICER measurements including J0437-4715 and J0614-3329 \cite{Vinciguerra_2024,Salmi_2024,Choudhury_2024,Mauviard_2025}. The 2D contours show the 90\% C.L.}
    \label{fig:correlation-plot-all-astro}
\end{figure}

In Fig.~\ref{fig:correlation-plot-all-astro}, we compare astrophysically constrained EoS posteriors, with and without the more recent constraints from J0437-4715 and J0614-3329, to the publicly available, astrophysical constrained EoS posterior of \citet{Legred_2021}. All results show broad consistency in both $R_{1.4}$ and $M_{\rm max}$, but some shifts are seen within the distributions of inferred parameters.

With the inclusion of GW170817, GW190425, and the constraints of the first two NICER pulsars J0030$+$0451 and J0740$+$6620 \cite{Miller_2019,Dittmann_2024}, we find that the new semiparametric samples infer radii with a distribution similar to  \citet{Legred_2021}; the limit on large radius comes from the reduced upper limit on pressure around nuclear saturation density from the meta-model constraints as discussed earlier for Fig.~\ref{fig:semiparam-gp-correlation}. The inferred radius at $1.4 M_{\odot}$ is $12.34_{-0.99}^{+0.70}$\;km for our semiparametric model after these observations. Our semiparametric distribution also continues to prefer higher $M_{\rm max}$, following from the broader prior range seen in Fig.~\ref{fig:semiparam-gp-correlation} before GWs and X-ray are included. Our inferred maximum mass with GWs, X-ray observations of J0030$+$0451 and J0740+6620 is $M_{\rm max} = 2.40_{-0.32}^{+0.45} M_{\odot}$. 

As the published results for X-ray observations of J0437-4715 and J0614-3329 favor smaller radii, similar to those inferred from GW170817, cumulative constraints from NICER observations of all four X-ray pulsars \cite{Vinciguerra_2024,Salmi_2024,Choudhury_2024,Mauviard_2025} drive the resulting EoS posterior towards a smaller radius at our reference mass. We report a median $R_{1.4}$ value of $11.44^{+0.98}_{-0.60}$\;km at 90\% C.L. The inclusion of these compact stars also restricts the maximum masses supportable by our semiparametric EoS family, and despite our broader prior we return to a maximum mass preference of $M_{\rm max} = 2.31_{-0.23}^{+0.35} M_{\odot}$.

\begin{figure}[t!]
    \centering
    \includegraphics[width=\columnwidth]{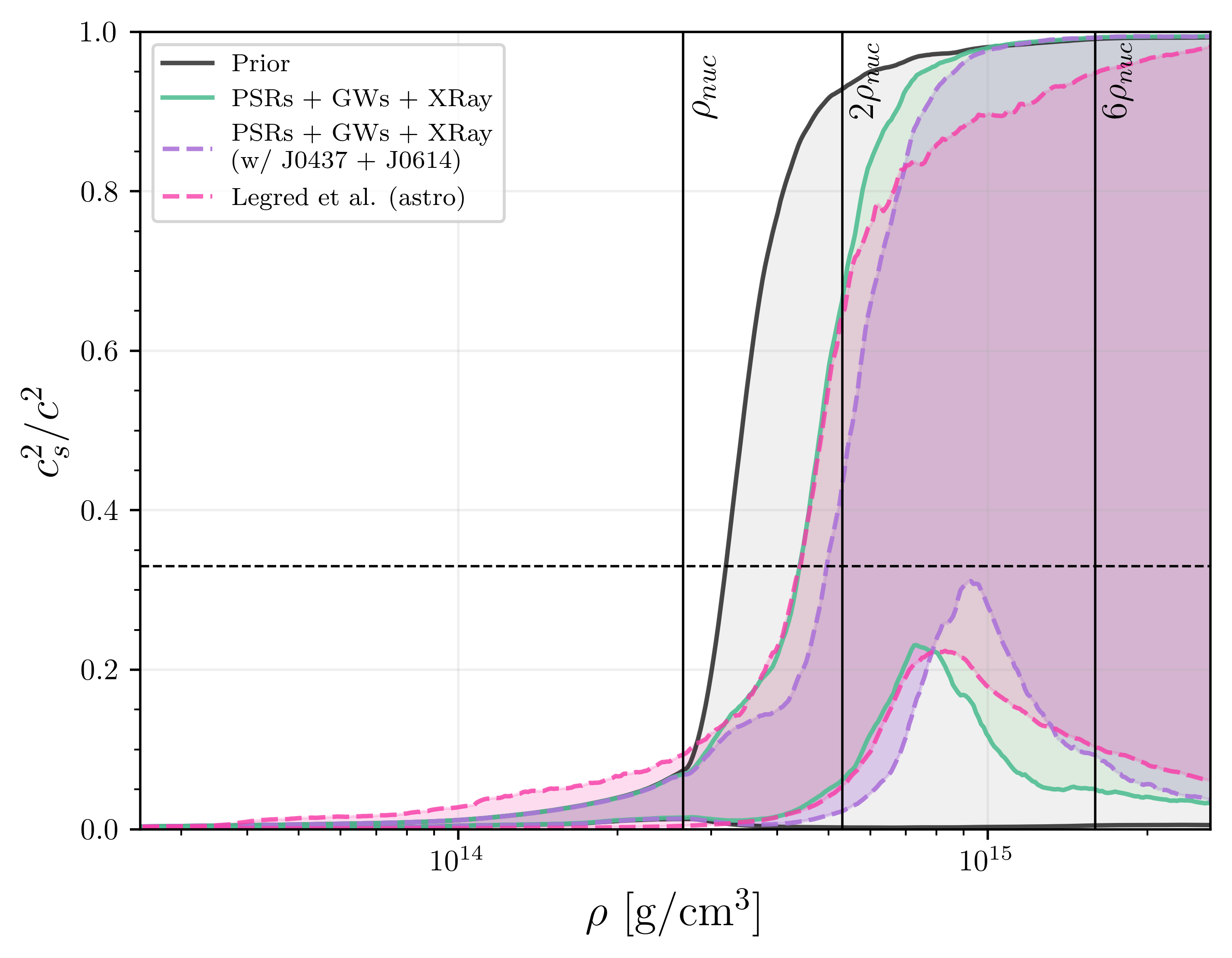}
    \caption{Posterior distributions of the speed of sound as a function of rest mass density, compared with the astrophysically constrained nonparametric EoS posterior of Legred et al. Pink lines correspond to that previous nonparametric posterior, and green lines represent our semiparametric EoS posterior. These use the NICER constraints of J0030+0451 and J0740+6620 \cite{Miller_2019,Dittmann_2024}. Purple contour lines denote our semiparametric EoS posterior with all public NICER measurements including J0437-4715 and J0614-3329 \cite{Vinciguerra_2024,Salmi_2024,Choudhury_2024,Mauviard_2025}. The 2D contours show the 90\% C.L.}    \label{fig:all-astro-sound-speed}
\end{figure}

The posterior distributions for the speed of sound are shown in Fig.~\ref{fig:all-astro-sound-speed}. Note that below nuclear saturation density, the semiparametric EoS are more constrained than the model-agnostic GP distribution due to the addition of nuclear information. When informed by the same PSRs + GWs + X-ray observations, the new semiparametric EoS posterior allows for a wider 90\% C.L upper and lower bound on $c_{s}^{2}$ above $2\rho_{\rm nuc}$ than the Legred et al. posterior distribution. We attribute this to our kernel and hyperparameter choices aimed at fully covering the pressure-density plane, as discussed in Section \ref{sec:methods/eos/GP}. 

At asymptotically higher densities ($\sim 40n_{\rm nuc}$) relevant to the regime of Quantum Chromodynamics (QCD), the speed of sound is expected to approach the conjectured conformal limit $c_{s}^{2}/c^{2} = 1/3$ (see Ref.~\cite{Bedaque_2015} for details).
However, there exist proposed theoretical models where $c_{s}^{2}/c^{2}$ rises above $1/3$ at intermediate densities~\cite{McLerran_2019,Kojo_2015}.  Previous works have found that current observations support a rise in sound speed $c_{s}^{2}/c^{2} \geq 1/3$ at intermediate densities relevant to NS matter~\cite{Tews_2018, Landry_2020, Altiparmak_2022, Fujimoto_2022, Cai:2023ajw}. 
Here, we similarly find that sound speeds less than $1/3$ have restricted support between 2 and $6\rho_{\rm nuc}$ with astrophysical inference using the semiparametric model, even as our GP framework has expanded the range of EoS support at high density.
When we include additional data from observations of PSR J0437-4715 and PSR J0614-3329, our posterior distribution is more constrained, with sound speed $\gtrsim 1/3$ at $\sim 4\rho_{\rm nuc}$ relative to the baseline semiparametric posterior.

Lastly, we note that the sound speed features introduced into the prior EoS set by training on models with phase transitions, as seen in Fig.~\ref{fig:changes-in-sound-speed}, are still present in draws from the posterior distribution and so are not excluded by current astrophysical observations.

\subsubsection{Comparison to constraints in other work}
\label{sec:results/mma/compare-constraints}

Our semiparametric EoS framework gives broadly consistent results with other work combining astrophysical constraints from current-generation observations, although some differences are seen to emerge from the choice of EoS prior. We have so far directly compared these results in Sec.~\ref{sec:results} with the previous GP results of Ref.~\cite{Legred_2021}, to highlight the impact of our restricted meta-model framework at low density and our new GP with hyperparameters chosen to cover a broad range of pressures at high density. We include in Table \ref{tab:astro_compare} a summary of results inferred from comparable recent astrophysical observations.
 
In the table, we briefly summarize the modeling used below and around nuclear saturation ($\sim \rho_\text{nuc}$) and in higher densities in the inner core ($\gtrsim \rho_\text{nuc}$); outer crust models may vary. Notation for models and observations follows usage of earlier Figures in this work where possible. 

For modeling, $\chi$ indicates Chiral Effective Field theory constraints, MM indicates the meta-model. Phenomenological descriptions of the EoS include piecewise polytropes in pressure-density (PP) , Gaussian Processes (GP), a parameterized description in sound speed (CS), and neural network models (NN). In contrast, some works use relativistic mean field (RMF) nuclear theory models.
 
Our baseline set of observations includes PSR J0348$+$0432 \cite{Antoniadis_2013}, PSR J0740$+$6620 \cite{Fonseca:2021wxt}, and PSR J1614$-$2230 \cite{Arzoumanian_2018}. The most informative mass measurement, that of PSR J0740$+$6620, is used by all works. GW denotes the use of GW170817 data~\cite{Abbott_2019_GW17properties} and optionally GW190425~\cite{Abbott_2020}. Our baseline X-ray uses NICER constraints of J0030+0451 and J0740+6620 \cite{Miller_2019,Dittmann_2024} and we compare to other results using those first two NICER X-ray pulsars.
 
Biswas et al.~\cite{Biswas_2024} uses a three-segment approach in density with a SLy representation of the EoS up to a junction point, nuclear empirical parameters up to $1$-$2\rho_{\rm nuc}$, and then piecewise polytropic extensions afterwards. Koehn et al.~\cite{Koehn_2025} uses a meta-model approach with a sound speed extension mechanism. Mauviard et al.~\cite{Mauviard_2025}, use piecewise polytropes with a sound speed high-density extension. Altiparmak et al. and Chimanski et al.~\cite{Altiparmak_2022, Chimanski_2023} use hybridized constructions of their EoSs, with a sound speed extension mechanism extending to asymptotically high densities. Char et al.~\cite{Char_2023} use a relativistic formulation of the meta-model. Malik et al.~\cite{Malik_2022} creates an EoS prior with a Relativistic Mean Field framework. Ref.~\cite{Li_2025} uses covariant density functional models. Ref.~\cite{Fan_2024} compares several different model frameworks, including a sound speed parameterization, a GP, and a neural network (NN). Several of these works also explore inclusion of additional observational data; for example Ref.~\cite{Malik_2022} includes mass measurements of PSR J1810+1714 and Ref.~\cite{Traversi_2020} additionally uses observations of quiescent low-mass X-ray binaries. Some of the above include constraints from perturbative Quantum Chromodynamic (pQCD) likelihoods in their analysis \cite{Fan_2024,Biswas_2024,Koehn_2025}. 
We also include some results including J0437-4715\cite{Vinciguerra_2024,Salmi_2024,Choudhury_2024} which leads to a moderately smaller radius inference\cite{Rutherford_2024, Biswas_2024}, and can also reduce the supported maximum mass. For details of the exact modeling and observational data used for each result, see the original works.

We emphasize that within the presented uncertainties, our results are consistent with all these existing inferences. Our $R_{1.4}$ posterior is very consistent with previous results that use similar astrophysical data. However, as noted in Sec.~\ref{sec:results/eos} and Sec.~\ref{sec:results/mma}, our GP includes more prior support for higher maximum masses. We find as a result that our preferred $M_\text{max}$, and our inferred upper bound on $M_\text{max}$, is larger than previous work.

In this work, we also demonstrate the impact of including both new NICER results, labeled as X-ray+J0437+J0614. This version of our posterior uses NICER measurements including both J0437-4715 and J0614-3329~\cite{Vinciguerra_2024,Salmi_2024,Choudhury_2024,Mauviard_2025}. We are again consistent with previous results using multiple high-density frameworks~\cite{Mauviard_2025}. However, we again prefer higher maximum mass, and somewhat smaller radius at 1.4 $M_{\odot}$, which we attribute to the broad coverage and flexibility of our high density GP.

\begin{table*}[t!]
    \centering
    \begin{tabular}{cccccc}
         & Model $\lesssim$ to $\sim \rho_\text{nuc}$ & Model $\gtrsim \rho_\text{nuc}$ & Observations & $R_{1.4}$\;(km) & $M_\textrm{max}$ ($M_\odot$ ) \\
        \hline \\
        This work & MM+$\chi$ & GP &  GWs + X-ray & $12.34_{-0.99}^{+0.70}$ & $2.40_{-0.32}^{+0.45}$ \\
        \hline \\ 
        Legred \textit{et al.} 2021~\cite{Legred_2021} & GP & GP & GWs + X-ray & $12.56^{+1.00}_{-1.07}$ & $2.21^{+0.31}_{-0.21}$ \\ 
        Altiparmak \textit{et al.} 2022~\cite{Altiparmak_2022} & $\chi$ & CS &  GWs + X-ray & $12.42_{-0.99}^{+0.52}$ & - \\
         Malik \textit{et al.}2022~\cite{Malik_2022} & RMF & RMF & GWs + X-ray + PSR J1810+1714 & $12.62_{-0.55}^{+0.59}$ & $2.144_{-0.123}^{+0.211}$ \\
         Char \textit{et al.} 2023~\cite{Char_2023} & Relativistic MM & Relativistic MM &  GWs & $12.72\pm0.46$ & - \\
         Fan \textit{et al.} 2024~\cite{Fan_2024} & FFNN/CS/GP & FFNN/CS/GP & GW + X-ray & - & $2.25_{-0.07}^{+0.08}$\\       
         Tsang \textit{et al.} 2024~\cite{Tsang_2024} & MM & MM & GWs + X-ray & $12.9_{-0.5}^{+0.4}$ & - \\
         Koehn \textit{et al.} 2025~\cite{Koehn_2025} & MM & CS & GWs + X-ray & $12.26_{-0.91}^{+0.80}$ & $2.25_{-0.22}^{+0.42}$ \\

        Rutherford \textit{et al.}~2024\cite{Rutherford_2024}  & PP + $\chi$ & PP & GWs + X-ray + J0437 & $12.30_{-1.04}^{+0.55}$ & $2.15\pm0.20$\\
         ``..." & PP + $\chi$ & CS & GWs + X-ray + J0437 & $12.29_{-1.03}^{+0.47}$ & $2.08_{-0.17}^{+0.25}$ \\
        Biswas \textit{et al.} 2024~\cite{Biswas_2024} & SLy & $\chi$ + PP & GWs + X-ray + J0437 & $12.34_{-0.53}^{+0.43}$ & $2.22_{-0.19}^{+0.21}$\\ 
        Li \textit{et al.} 2025~\cite{Li_2025} & CDF & CDF & GWs + X-ray + HESS J1231-1411 & $12.47_{-0.50}^{+0.48}$ & $2.20_{-0.17}^{+0.23}$\\
        \hline \\
        This work & MM+$\chi$ & GP  & GWs + X-ray + J0437+J0614 & $11.44^{+0.98}_{-0.60}$ &$2.31_{-0.23}^{+0.35}$ \\
                \hline \\

        Mauviard \textit{et al.} 2025~\cite{Mauviard_2025} & PP + $\chi$ & PP  & GWs + X-ray+J0437+J0614 &  $12.05^{+0.56}_{-0.79}$ & - \\
        ``...'' & PP + $\chi$ & CS & GWs + X-ray+J0437+J0614 &$11.71^{+0.71}_{-0.63}$ & - \\

    \end{tabular}
    \caption{A comparison of our reference astrophysical quantities to a range of recent works where the same quantities inferred with a variety of modeling frameworks and observational data. The most massive well-measured PSR J0740+6620 is used in all results. Notation follows previous figures when possible; see text for details. Ranges presented follow selected summary statements in the references and may vary in the percentile covered (e.g. 90 or 95).}
    \label{tab:astro_compare}
\end{table*}

\section{Conclusions}
\label{sec:conclusion}

In this work, we construct a semiparametric framework for generating an EoS prior that utilizes both nuclear physics constraints, and model agnostic GP correlation structures to probe the properties of dense matter as learned from observations of NS's. We examine the imprint of the training data in our model-informed GP and find that our semiparametric EoS prior emulates sound-speed features of the training EoS distribution. With the minimal constraints of causality and supporting observations of heavy pulsars, we compare the semiparametric EoS framework to fully non-parametric EoS with a more model-agnostic GP representation, and meta-model EoS with piecewise polytropic extensions at high density. We find that using the Meta-model with the inclusion of $\chi$EFT constraints on the pressure up to the stitching point of $\rho_{\rm tr} = 150.2$\;MeVfm$^{-3}$ ($2.65\times10^{14}$ g/cm$^3$) gives significantly tighter posteriors on the EoS at sub-saturation densities. As seen in Fig.~\ref{fig:semiparam-gp-correlation}, this generates an upper bound on allowed radii with minimal constraints, excluding values above $\sim 14$\;km at $1.4 M_{\odot}$. However, our GP is designed to more densely sample higher pressure EoS that are still compatible with causality, so explores a wider range of maximum masses, extending past $3.2 M_{\odot}$. When compared to the Meta-model with piecewise polytropic extensions in the pressure-density space, the semiparametric model explores a wider range of pressures than the piecewise polytropes, corresponding to a larger range in radii after imposing causality and requiring support of massive pulsars.

Finally, we demonstrate astrophysical inference of the dense matter EoS with the semiparametric framework. With the inclusion of two recent NICER millisecond pulsar analyses of J0437-4715 and J0614-3329, we report a median value for $R_{1.4} = 11.44^{+0.98}_{-0.60}$\;km at the 90\% credible level.  The smaller inferred radii with these new observations implies that neutron star-black hole gravitational-wave events, such as those observed by LIGO, Virgo, and KAGRA \cite{LIGOScientific:2021qlt,LIGOScientific:2024elc}, may produce less ejected mass than previously inferred.

The maximum mass is used to classify gravitational-wave events without EM counterparts or strong tidal constraints \cite{2020ApJ...904...80E}. Without the X-ray timing results of J0437-4715 and J0614-3329, our EoS distribution can allow stable neutron stars up to a mass of $2.8 M_{\odot}$. Notably, this would support the interpretation of the $2.6 M_{\odot}$ component of GW190814 as more likely a neutron star compared to previous assessment \cite{LIGOScientific:2020zkf}.
Including X-ray timing results from J0437-4715 and J0614-3329 moves the classification threshold toward previous GW-only assessments, as we find $M_{\rm max} = 2.31_{-0.23}^{+0.35} M_{\odot}$. However, our demonstration that causal EoS with realistic crusts can allow support for more massive stars can inform how future GW observations are interpreted with minimal assumptions.

The comprehensive coverage of the EoS space that we demonstrate in this work will become increasingly important as we move towards high precision measurements with future nuclear experiment, radio, X-ray, and gravitational-wave astronomy facilities. To avoid decimation of a fixed-sample prior during astrophysical inference (as expected from higher precision measurements from next-generation facilities), we can build on our code development to allow inference to ``sample on the fly'' from the GP as seen for example in Ref.~\cite{Gong:2024lhq}. New constraints could then directly update where in the parameter space the EoS distribution should be drawn from. However, for near-term use of our semiparametric model in interpreting astronomical observations, we provide publicly available sets of EoS that reflect the distributions explored in this work as accompanying data.

\section{Acknowledgments}
The authors thank Valerie Poynor, the developers of CUTER, Anthea Fantina and Francesca Gulminelli for useful discussions. 
S.N, L.T, and J.R acknowledges support by NSF grants PHY-2409736, PHY-2110441, and the Nicholas and Lee Begovich Center for Gravitational-Wave Physics and Astronomy. J.R. acknowledges support from Perimeter Institute. I.L. acknowledges support from the DOE under award number DE-SC0023101.
L.S acknowledges the financial support of the National Science Foundation Grant No. PHY 21-16686.
The authors are grateful for computational resources provided by the LIGO Laboratory and supported by National Science Foundation Grants PHY-0757058 and PHY-0823459.
This material is based upon work supported by NSF's LIGO Laboratory which is a major facility fully funded by the National Science Foundation.
This research has made use of data or software obtained from the Gravitational Wave Open Science Center (gwosc.org), a service of the LIGO Scientific Collaboration, the Virgo Collaboration, and KAGRA. This material is based upon work supported by NSF's LIGO Laboratory which is a major facility fully funded by the National Science Foundation, as well as the Science and Technology Facilities Council (STFC) of the United Kingdom, the Max-Planck-Society (MPS), and the State of Niedersachsen/Germany for support of the construction of Advanced LIGO and construction and operation of the GEO600 detector. Additional support for Advanced LIGO was provided by the Australian Research Council. Virgo is funded, through the European Gravitational Observatory (EGO), by the French Centre National de Recherche Scientifique (CNRS), the Italian Istituto Nazionale di Fisica Nucleare (INFN) and the Dutch Nikhef, with contributions by institutions from Belgium, Germany, Greece, Hungary, Ireland, Japan, Monaco, Poland, Portugal, Spain. KAGRA is supported by Ministry of Education, Culture, Sports, Science and Technology (MEXT), Japan Society for the Promotion of Science (JSPS) in Japan; National Research Foundation (NRF) and Ministry of Science and ICT (MSIT) in Korea; Academia Sinica (AS) and National Science and Technology Council (NSTC) in Taiwan.

\section{Data Availability}

The finalized data that supports the findings of this study is publicly available at URL/DOI: \href{https://doi.org/10.5281/zenodo.15801145}{10.5281/zenodo.15801144}.

\appendix
\section{Hyperparameter Choices}
\label{sec:appendix/hyperparameters}

\begin{figure*}[t]
    \begin{subfigure}[t]{0.33\hsize}
    \resizebox{\hsize}{!}{\includegraphics{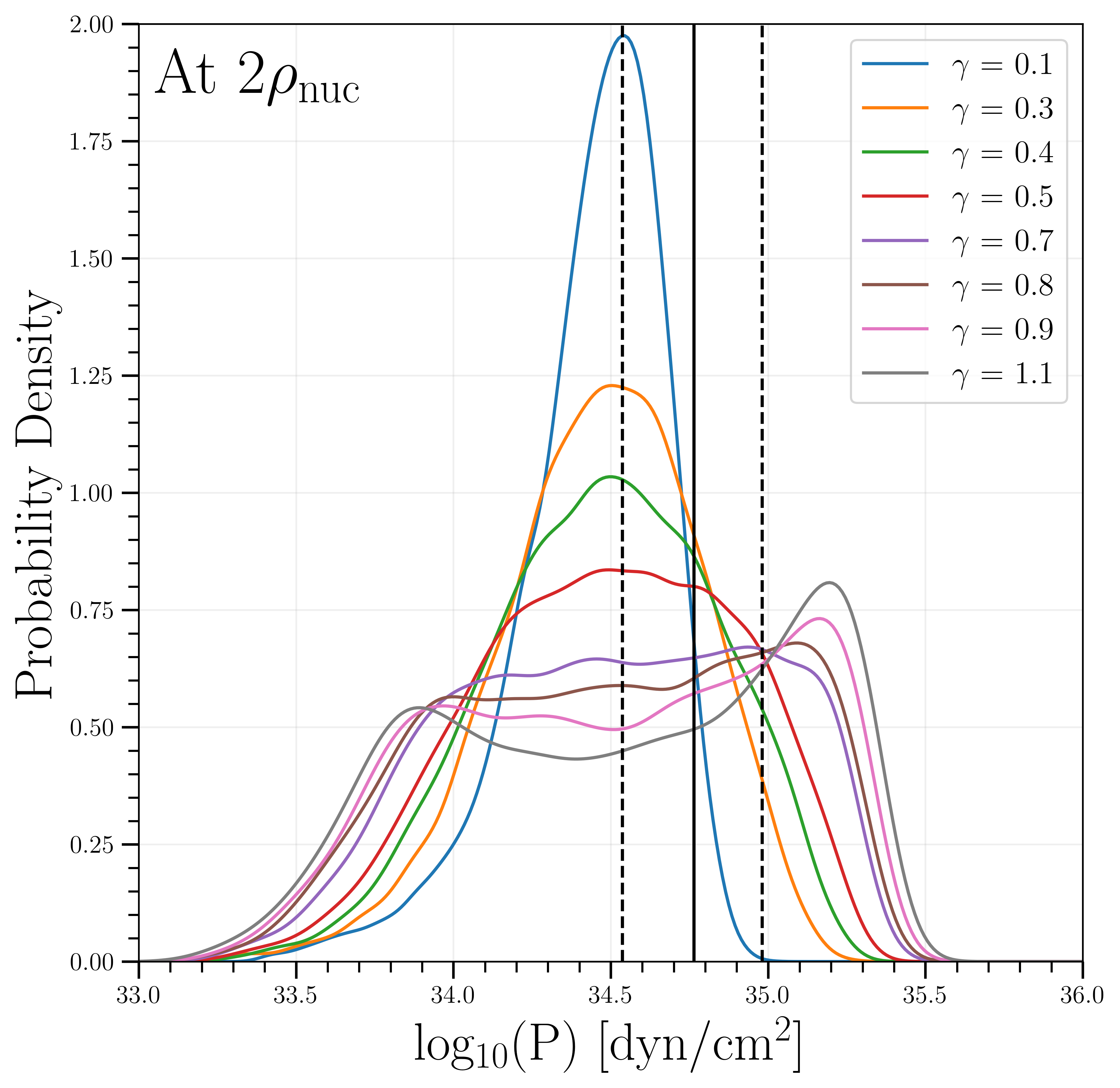}}
    \end{subfigure}%
    \begin{subfigure}[t]{0.33\hsize}
    \resizebox{\hsize}{!}{\includegraphics{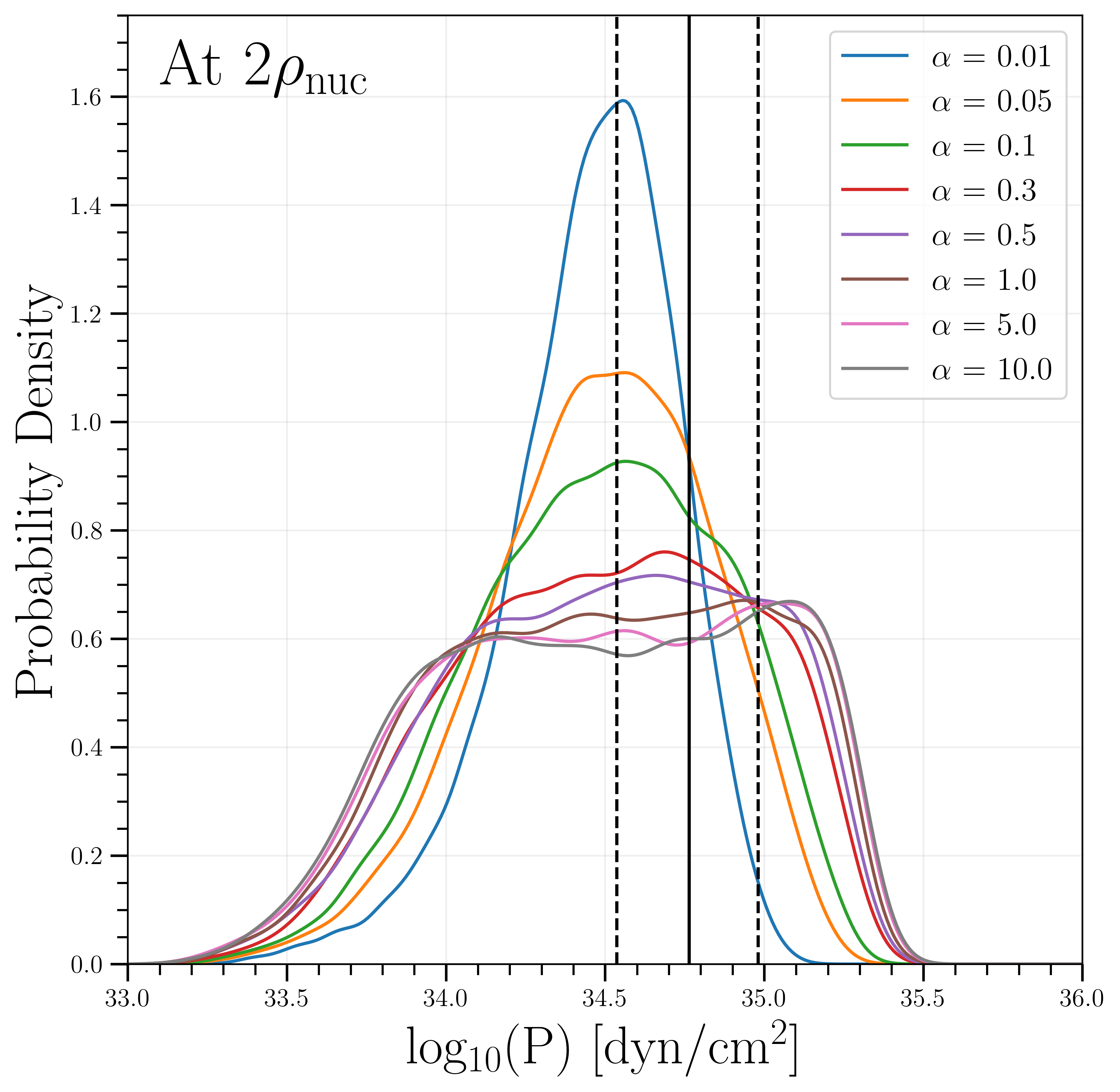}}
    \end{subfigure}%
    \begin{subfigure}[t]{0.33\hsize}
    \resizebox{\hsize}{!}{\includegraphics{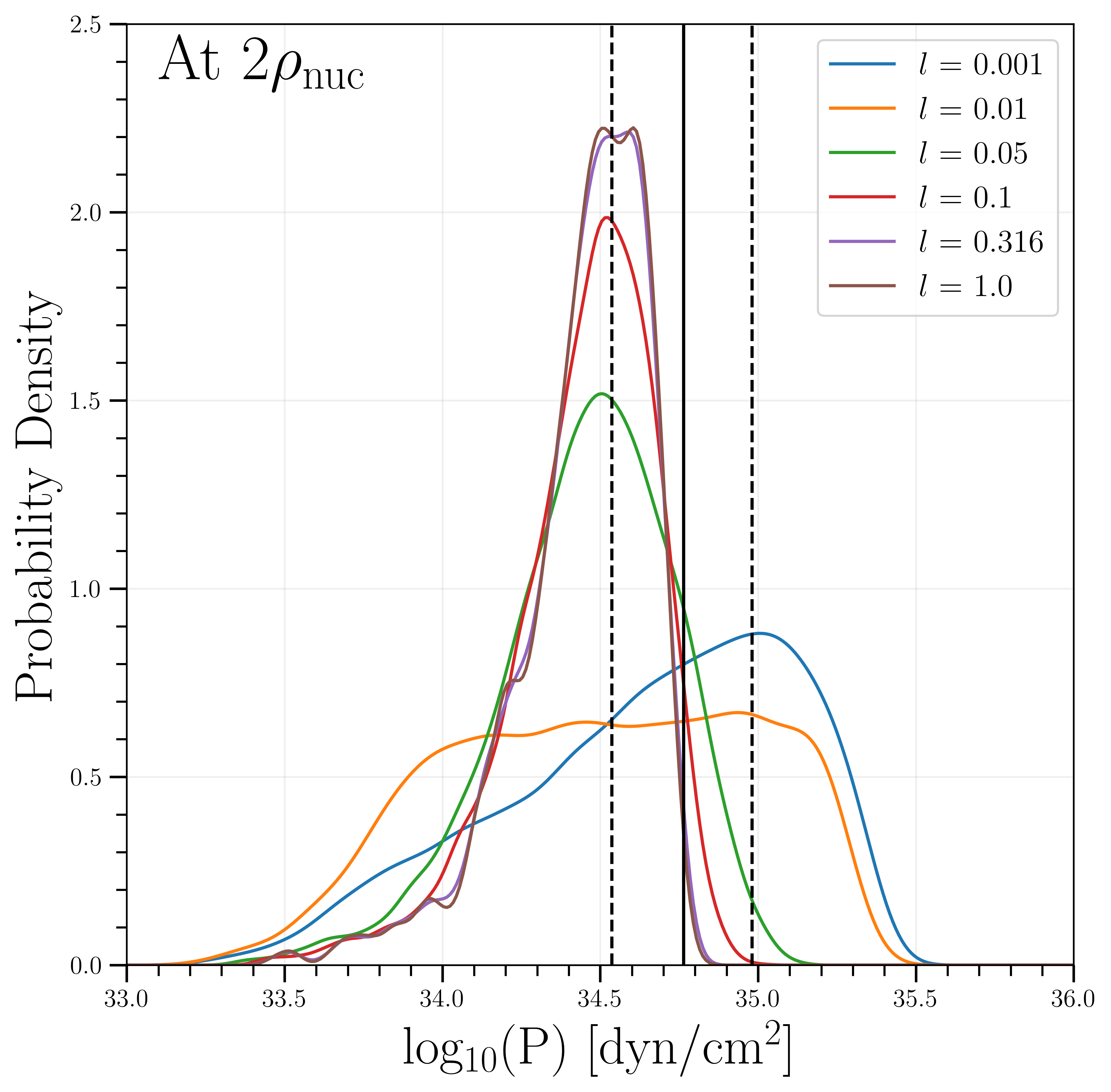}}
    \end{subfigure}%
    
    \vspace{0.05cm}

    \begin{subfigure}[b]{0.33\hsize}
    \resizebox{\hsize}{!}{\includegraphics{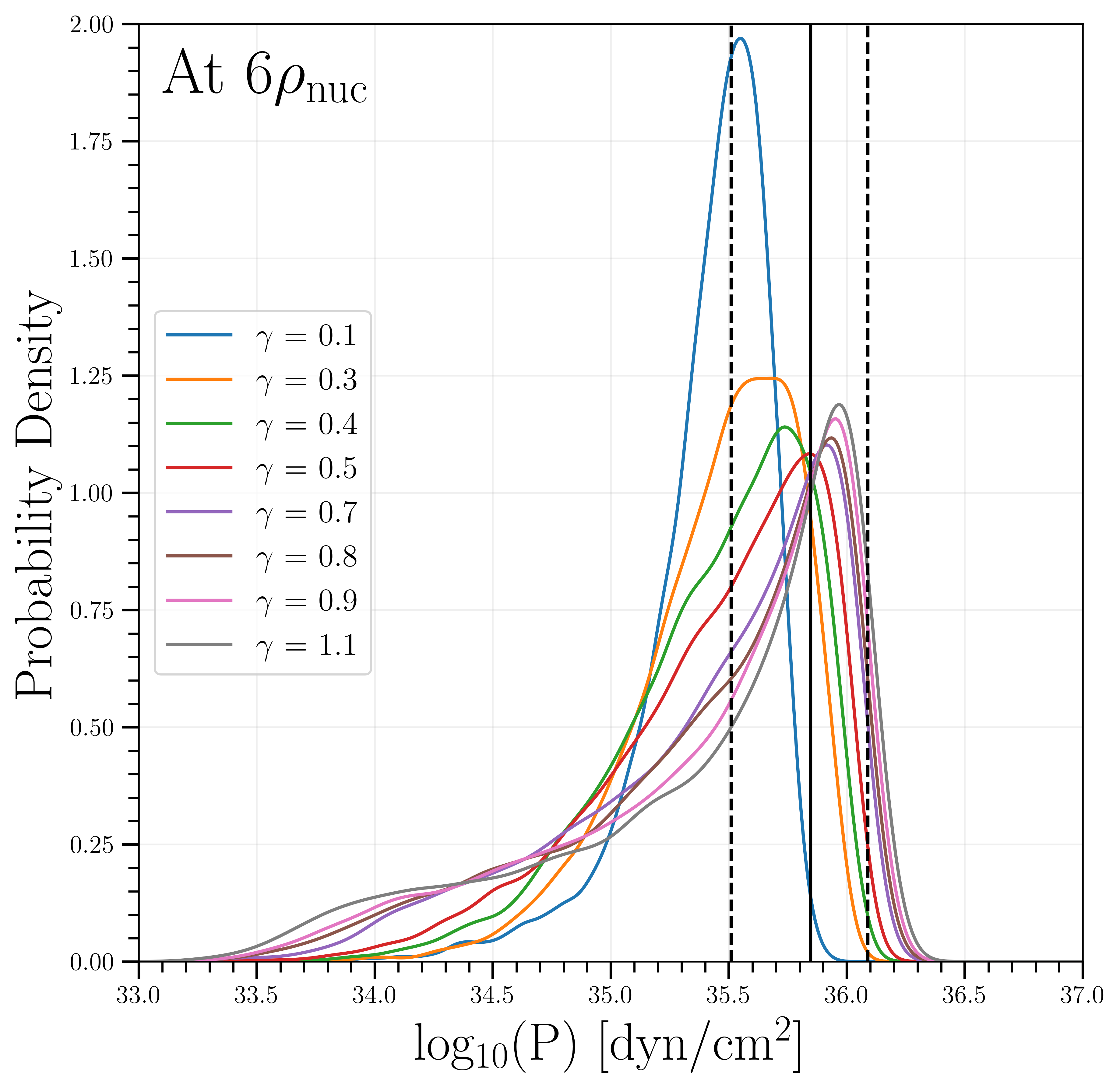}}
    \end{subfigure}%
    \hfill
    \begin{subfigure}[b]{0.33\hsize}
    \resizebox{\hsize}{!}{\includegraphics{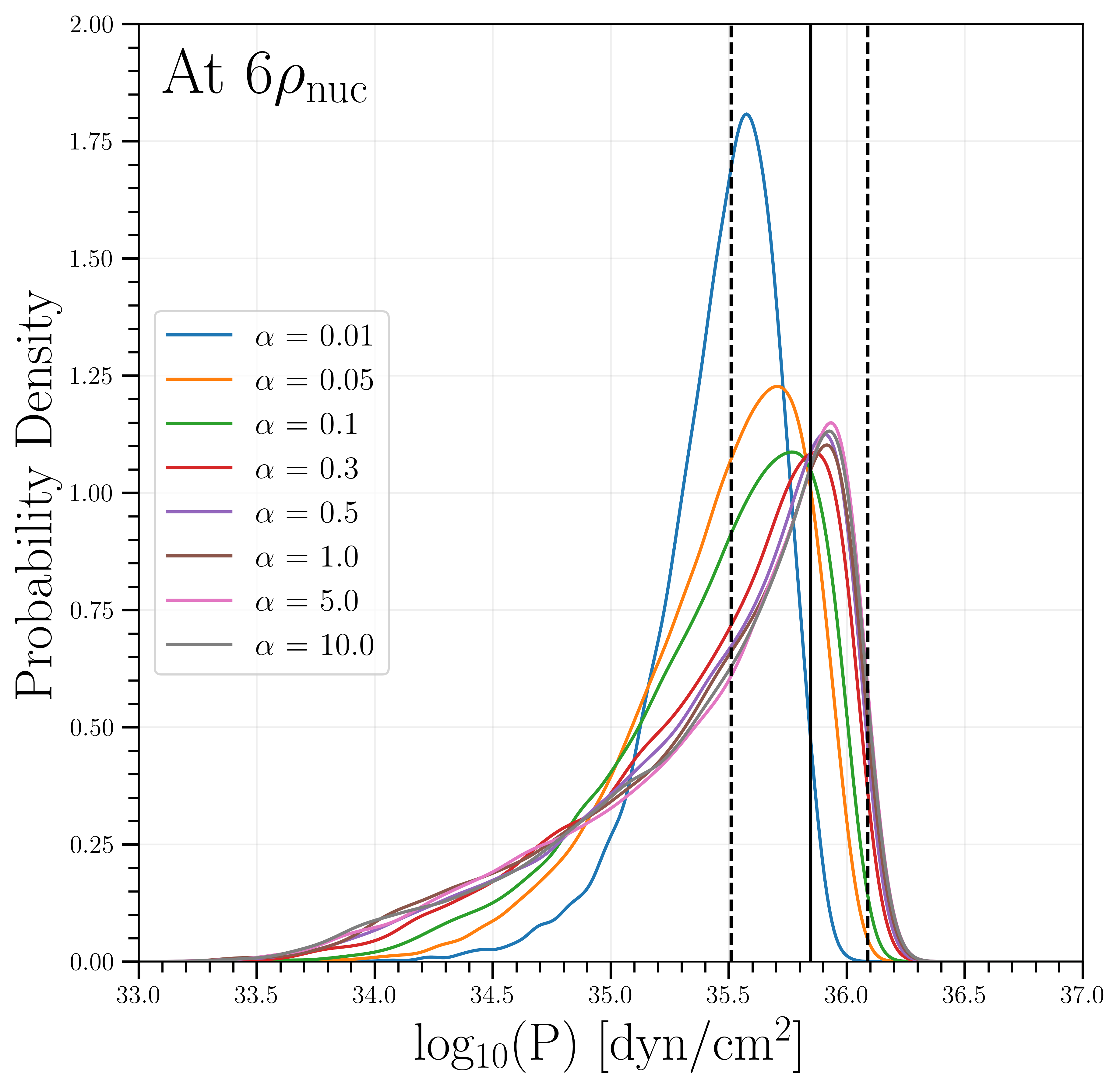}}
    \end{subfigure}%
    \begin{subfigure}[b]{0.33\hsize}
    \resizebox{\hsize}{!}{\includegraphics{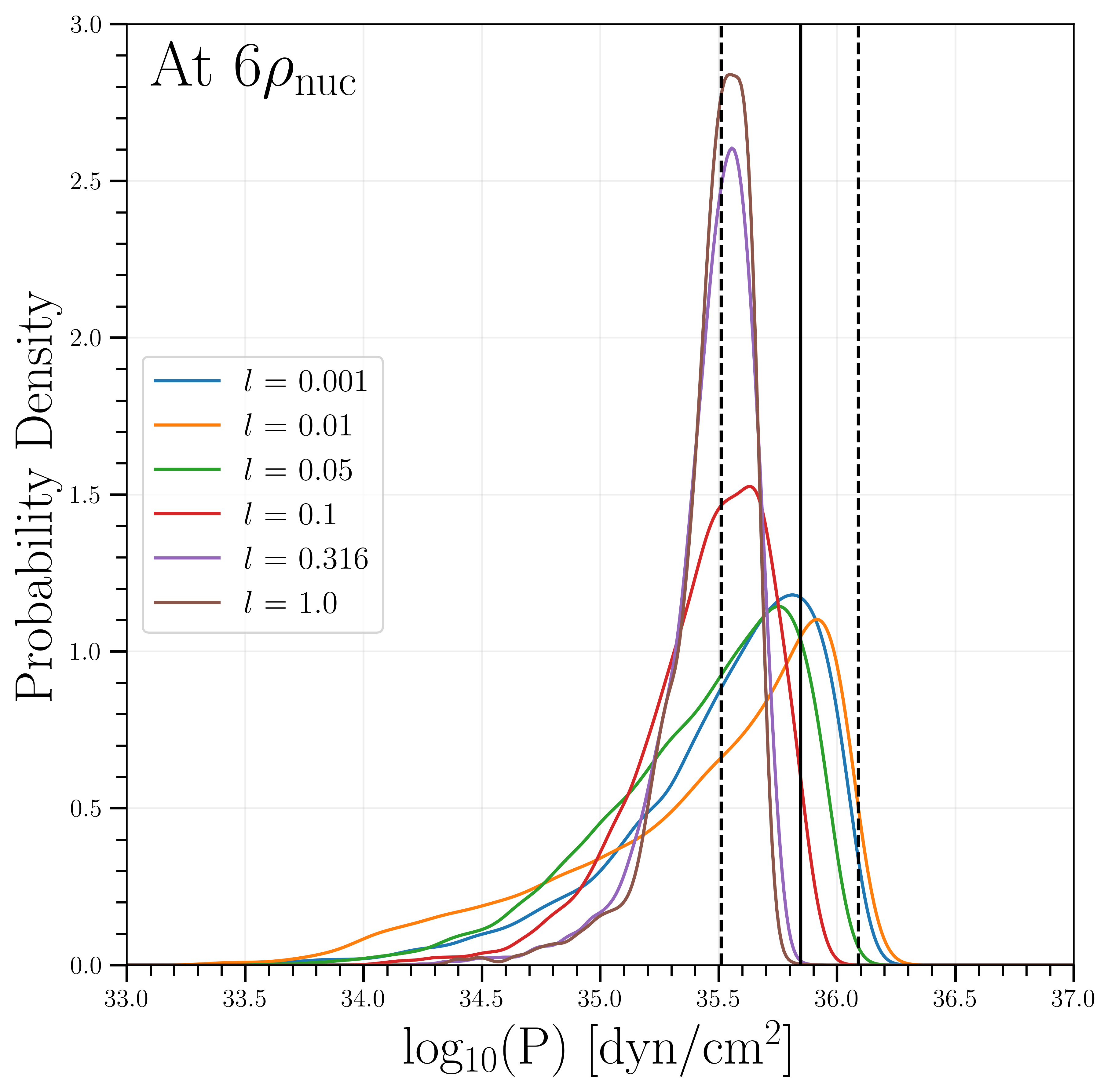}}
    \end{subfigure}%
    \caption{Probability density function of pressure values found at $2\rho_{\rm nuc}$ and $6\rho_{\rm nuc}$, illustrating effects of hyperparameter choice. The black vertical line represents the median pressure of the nuclear training data at the respective fixed densities, whereas the dashed black lines are the 90\% credible level bounds.}
    \label{fig:hyperparameter-control}
\end{figure*}

The impact of our choices in constructing the semiparametric EoS prior model can be seen by varying the hyperparameters of the rational quadratic kernel, and verifying their effects in the pressure-density space. In Fig.~\ref{fig:hyperparameter-control}, we display the coverage of the EoS distribution in terms of log pressure at fixed densities, namely $2\rho_{\rm nuc}$ and $6\rho_{\rm nuc}$, when altering the covariance strength $\gamma$, the scale mixture $\alpha$, and the correlation length scale $\ell$, with control values of $\gamma =. 7$, $\alpha = 1.0$, and $\ell = 0.01$. Beginning with the prior at $2\rho_{\rm nuc}$, smaller values of the covariance strength and scale mixture caused peaked distributions in the pressure values at $\sim 10^{34.5}$ [dyn/cm$^{2}$]. With the covariance strength set towards lower values, the assumed variability described within the length scales are not enforced as strongly, and therefore yield narrower coverage in the pressure. Meanwhile, with lower $\alpha$ values, we see a similar modal structure in the pressure coverage. As $\alpha \rightarrow 0$, the variations on small and large scales become smaller and highly correlate points across all distances, thus also yielding a narrower pressure distribution due to information from low density. Most prominently, changing the correlation length scale $\ell$ towards lower values has an inverse effect from lowering $\gamma$ and $\alpha$, yielding a broader EoS distribution at both reference densities. 

For $\gamma < 0.7$, the distribution exhibits a single prominent peak, though it begins to flatten progressively as the hyperparameter approaches 0.7. As we increase $\gamma$, the distribution begins to dip at the center until $\gamma = 1.1$ and the distribution has developed a bimodal structure at around $10^{33.8}$ and $10^{35.3}$ [dyn/cm$^{2}$]. We can look at the effects of $\alpha$ in terms of orders of magnitude from 0.01 to 10. Significant changes occur over the first three orders of magnitude in $\alpha$ but beyond $\alpha = 1$, the effect seemingly plateaus, indicating a possible saturation point in its influence on the distribution’s shape. At $6\rho_{\rm nuc}$, the effects of varying $\gamma$ and $\alpha$ appear to be very similar to one another, with comparable structure in the coverage of pressures and their modalities. Larger $\gamma$ and $\alpha$ values result in strong, left-skewed distributions peaking around $10^{36}$ [dyn/cm$^{2}$]. As these hyperparameters decrease, both distributions demonstrate a shift towards a median pressure value of $\sim 10^{35.5}$ [dyn/cm$^{2}$] overlapping with the lower bound of the 90\% confidence interval of the nuclear training data. Note that although the plots in Fig.~\ref{fig:hyperparameter-control} are displayed in logarithmic scale, in linear space we obtain a relatively flat distribution. This is important since we want our prior to be uniformly distributed such that we may resolve the parameter space well.

\section{EoS Construction Details}
\label{sec:appendix/prior-construction}

In practice, one can arbitrarily stitch EoS's at any given density. We emphasize that the transition density to stitch extensions onto is ultimately a choice. Higher stitching densities, and therefore employing $\chi$EFT and nuclear experimental constraints up to larger densities, can lead to tighter constraints on the pressure values and astrophysical parameters. We have compared transition densities between 1 and 1.25$\rho_{\rm nuc}$ and our studies found modest impact in regards to limits on NS radii, favoring lower radii values and smaller $M_{\rm max}$. Some works, such as Refs.~\cite{Essick:2020flb,Rutherford_2024} have used astronomical data to specifically constrain the transition from nuclear modeling to high-density flexibility; our formalism gives broadly consistent results and could also be extended to marginalize over transition densities to reproduce these constraints in the transition region.

While constructing our EoS prior, we studied the effects of low density variability of the EoS at the stitching point as opposed to stitching GP extensions onto a singular fixed crust. Taking into account that our implementation of GP extensions are model-informed, we find that the resulting distribution of EoS using a singular fixed crust had a reduced range in the pressure-density space, excluding lower pressures and strong phase transition like features, despite using the same hyperparameter configuration and training data set. When we allow for informed uncertainty in the low density meta-model EoS, the set of per-meta-model higher-density GP act as a Gaussian mixture model, broadening the range of pressure-density explored. As causality and thermodynamic consistency limit how the GP extends to higher densities, information propagated from the uncertainty bounds from $\chi$EFT and nuclear experimental measurements therefore also informs how the global GP varies towards higher densities.

\bibliography{References}

\end{document}